\documentclass{aa}
\usepackage{graphics,epsfig,amsmath,amssymb,amstext,txfonts}

\def\la{\mathrel{\mathchoice {\vcenter{\offinterlineskip\halign{\hfil
$\displaystyle##$\hfil\cr<\cr\sim\cr}}}
{\vcenter{\offinterlineskip\halign{\hfil$\textstyle##$\hfil\cr
<\cr\sim\cr}}}
{\vcenter{\offinterlineskip\halign{\hfil$\scriptstyle##$\hfil\cr
<\cr\sim\cr}}}
{\vcenter{\offinterlineskip\halign{\hfil$\scriptscriptstyle##$\hfil\cr
<\cr\sim\cr}}}}}

\begin{document}


\title{Propagation of high-energy cosmic rays in extragalactic turbulent magnetic fields: resulting energy spectrum and composition}

\author{No\'emie Globus, Denis Allard and Etienne Parizot}

\institute{Laboratoire Astroparticule et Cosmologie (APC), Universit\'e Paris 7/CNRS, 10 rue A. Domon et L. Duquet, 75205 Paris Cedex 13, France.}

\offprints{denis.allard@apc.univ-paris7.fr}

\date{Received date; accepted date}

\abstract{We extend previous studies of mixed-composition extragalactic cosmic-ray source models, by investigating the influence of a non-negligible extragalactic magnetic field on the propagated cosmic-ray spectrum and composition. We study the transport of charged particles in turbulent fields and the transition from a ballistic to a diffusive propagation regime. We introduce a method allowing a fast integration of the particle trajectories, which allows us to calculate extragalactic cosmic-ray spectra in the general case, without using either the diffusive or the rectilinear approximation. We find that the main features of the mixed-composition models -- regarding the interpretation of the ankle and the non-monotonous evolution of the average cosmic-ray mass -- remain essentially unchanged as long as the magnetic field intensity does not exceed a few~nG.
}

\authorrunning{N. Globus, D. Allard, and E. Parizot}
\titlerunning{Propagation of UHE cosmic-ray nuclei in extragalactic magnetic fields}

\maketitle

\section{Introduction}
\label{Introduction}
The transition from galactic to extragalactic cosmic-rays is currently one of the most debated issues in high energy cosmic-ray physics. On the one hand, the KASCADE experiment (Antoni et al., 2005) has shown that the composition is getting heavier between the energy of the knee ($\sim 4\,10^{15}$ eV) and $10^{17}$ eV. The most likely interpretation of this observation is that the different elemental components of the galactic cosmic-rays experience successive drops (at energies scaling either with the charge or the mass) in this energy range. Although this conclusion seems to be independent of the hadronic model assumed, the details of the evolution of the different components remain unclear, which prevents, so far, KASCADE observations from putting strong constrains on the exact composition of the galactic component and its evolution in this energy range and above. On the other hand, experiments at higher energies (HiRes-Mia: Abu-Zayyad et al., 2000; Fly's eye: Bird et al., 1993; Yakutsk: Knurenko et al., 2006; Akeno: Dawson et al., 1998; Shinozaki et al., 2004; Haverah Park: Ave et al., 2003; HiRes: Abbasi et al., 2005) seem to favor a composition getting lighter between $10^{17}$~eV and $10^{18}$~eV and dominated by light elements above $10^{18}$ eV. This picture suggests that the transition from a heavy galactic component  to a light extragalactic component occurs somewhere between $10^{17}$ and $10^{19}$ eV. It is however difficult to better determine the energy scale and the structure of the transition from these experiments (based either on ground array or fluorescence techniques). As a result, several scenarios for the transition from galactic cosmic-rays (GCRs) to extragalactic cosmic-rays (EGCRs) are still viable.
In Allard et al. (2005, 2007a, 2007b), we proposed to explain the current high energy observations with an extragalactic (proton-dominated) mixed composition and showed that in this case, the ankle of the cosmic-ray spectrum can be interpreted as the signature of the transition from GCRs to EGCRs. We also showed that the resulting features in the energy evolution of the average maximum of longitudinal development of cosmic-ray air showers, $\langle X_{\max} \rangle$, are the most compatible with current experimental data. These studies were made assuming negligible extragalactic magnetic fields.

The extragalactic magnetic fields (EGMF) are currently poorly constrained, and often assumed to be negligible. Indeed, no convincing mechanism that could create strong magnetic fields over very large scales has yet been proposed. However, the current  upper limits do not exclude the presence of magnetic fields of several nG in the extragalactic medium (Kronberg, 1994; Blasi et al., 1999). In the case of non negligible magnetic fields, several effects on the predicted spectrum can be expected as a result of the existence of magnetic horizons (Stanev et al., 2000) evolving with the rigidity of charged particles. Strong magnetic fields, for instance, have been invoked to explain the absence of the GZK cut-off (Greisen, 1966; Zatsetpin and Kuzmin, 1966) claimed by the AGASA collaboration (see Aloisio and Berezinsky, 2004; Deligny et al., 2004). However, a sharp decrease of the cosmic-ray flux above $\sim~5\,10^{19}$~eV is observed by HiRes (Thomson et al., 2006) and now with larger statistics by Auger (Yamamoto et al., 2007).

\begin{figure*}[ht]
\centering{\includegraphics[width=\textwidth]{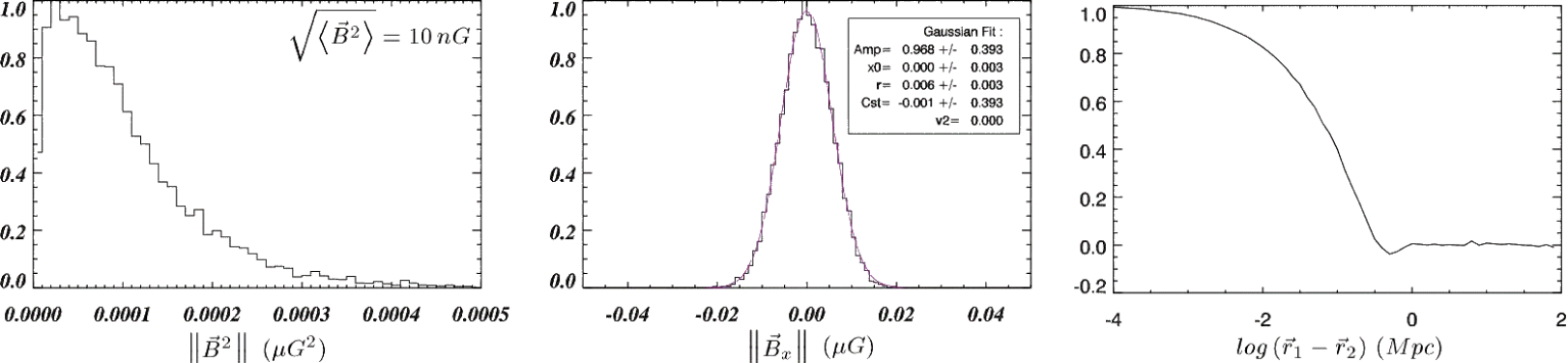}}
\caption{Structure of a purely turbulent magnetic field with variance $\sqrt{\langle \|\mathbf{\vec{B}^2}\| \rangle} = 10$~nG and maximum turbulence scale $\lambda_{\mathrm{max}} = 1$~Mpc. From left to right: distribution of the magnetic field norm squared, $\|\mathbf{\vec{B}^2}\|$, distribution of component $B_x$, and magnetic field two-point autocorrelation as a function of point separation.}
\label{fig:fieldStructure}
\end{figure*}

In the context of the  \emph{second-knee transition model} (Berezinski et al., 2002;  Aloisio et al., 2006), it has been shown that magnetic fields of the order of a nG could produce a suppression of the extragalactic component at low energy -- say below $\sim 10^{18}$~eV (Lemoine, 2005;  Aloisio et al., 2005; Kotera and Lemoine, 2007; Berezinsky and Gazizov, 2007) -- due to the existence of a magnetic horizon caused by the combined effect of the magnetic field and the source granularity (see below). The importance of this \emph{low energy cut} mechanism, originally introduced to limit the predicted extragalactic protons abundance at low energy (Lemoine, 2005), is a matter of debate. In the context of the second-knee transition model, it results in a very abrupt change of composition between $10^{17}$~eV and $10^{18}$~eV that would induce a very steep $\langle X_{\max} \rangle$ evolution with energy, possibly in conflict with current observations (see discussions in Allard et al., 2007, 2007b). Nevertheless, these studies show that non-negligible EGMFs could modify the UHECR spectrum.

In the present paper, we investigate the impact of non-negligible EGMFs on the observed cosmic-ray spectrum and composition in the case of an extragalactic mixed composition. In the next section, we present our modeling of extragalactic turbulent magnetic fields and particle transport. We then study the different transport regimes of energetic charged particles in turbulent magnetic fields. We finally calculate the extragalactic spectrum and the energy evolution of $\langle X_{\max} \rangle$ for different combinations of magnetic fields, source density and source power evolution with redshift, and discuss our results.

\section{The modeling of magnetic fields} 

Extragalactic magnetic fields are poorly known. Their spatial distribution, intensity, coherence length, time evolution and origin are uncertain. The presence of $\mu$G fields in the core of large galaxy clusters seems to be favored by the observations. However, the spatial extension of these large field zones and their volume filling factor in the universe are difficult to evaluate. Recent efforts have been made to model local magnetic fields using hydrodynamic simulations (e.g. Dolag et al., 2002; Miniati and Sigl, 2004; Kang et al., 2007), but their results sometimes differ by orders of magnitude. The resulting deflections expected for $10^{20}$~eV protons are thus radically different under different assumptions and lead to opposite conclusions about whether charged particle astronomy is possible or not at ultra-high energies. Even though the simulations agree that the magnetic field should be very weak in the large inter-cluster voids ($B \la 10^{-11}$~G), it is important to note that some potential sources of magnetic fields in the extragalactic medium are omitted in these simulations (see Parizot, 2004; Kotera and Lemoine, 2007 for discussions and references therein). In contrast, an interesting simple alternative to complex hydrodynamical simulations, offering more freedom to test different models of magnetic field evolution with the local density, has been proposed recently by Kotera and Lemoine (2007).

In view of the above-mentioned uncertainties, we shall use here a simplified approach, assuming that the universe is filled with a purely turbulent, homogeneous magnetic field, and will study the dependance of the high-energy cosmic-ray spectrum and composition on several physical parameters such as the variance of the turbulent field, the source density and the source power evolution.
 
To analyze the different propagation regimes of the charged particles in a purely turbulent field, we have performed a numerical simulation of the particle trajectories. The purely turbulent field is  represented by a sum of $N_{\mathrm{m}}$ modes as in Giacalone and Jokipii (1999) (see Casse et al., 2002; Deligny et al., 2004, for an other method based on Fast Fourier Transform):
\begin{equation}
B = \sum_{1}^{N_{\mathrm{m}}}A_{k_{n}}\hat\xi_{n}\exp(ik_{n}z^\prime_{n} + i\beta_{n}),
\label{eq:BGC}
\end{equation}
where $\hat\xi_{n} = \cos\alpha_{n}\hat{\mathbf{x}}^\prime_{n} + i \sin\alpha_{n}\hat{\mathbf{y}}^\prime_{n}$, $\alpha_{n}$ and $\beta_{n}$ are random phases (chosen once for all) and $[x^\prime_{n},y^\prime_{n},z^\prime_{n}] = [\mathcal{R}(\theta_{n},\phi_{n})]\times[x,y,z]$ are coordinates obtained by a rotation of the reference frame bringing the z axis in the direction of the $n^\mathrm{th}$ contributing wave (i.e. $\mathbf{k}_{n}$ is in direction $[\theta_{n},\phi_{n}]$, also chosen randomly). The amplitude $A(k_{n})$ is determined as a function of $\|\mathbf{k}_{n}\|$ according to a specific assumption on the type of turbulence (here, we assume a Kolmogorov spectrum with wavelengths between $\lambda_{\mathrm{min}}$ and $\lambda_{\mathrm{max}}$) and the chosen field variance (see Giacalone and Jokipii, 1999, for more details).

\begin{figure*}[ht]
\centering{\includegraphics[width=0.24\textwidth]{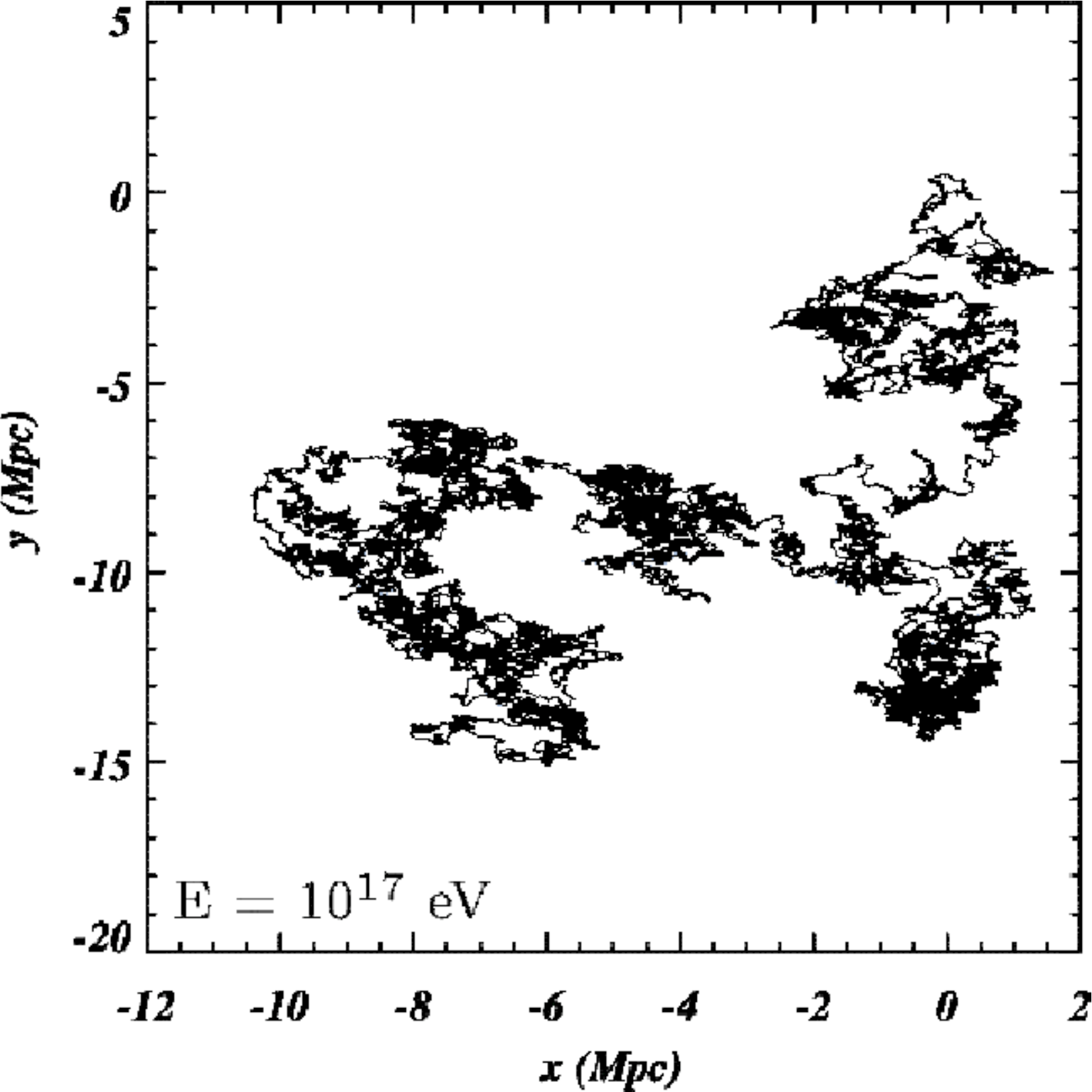}\hfill
\includegraphics[width=0.24\textwidth]{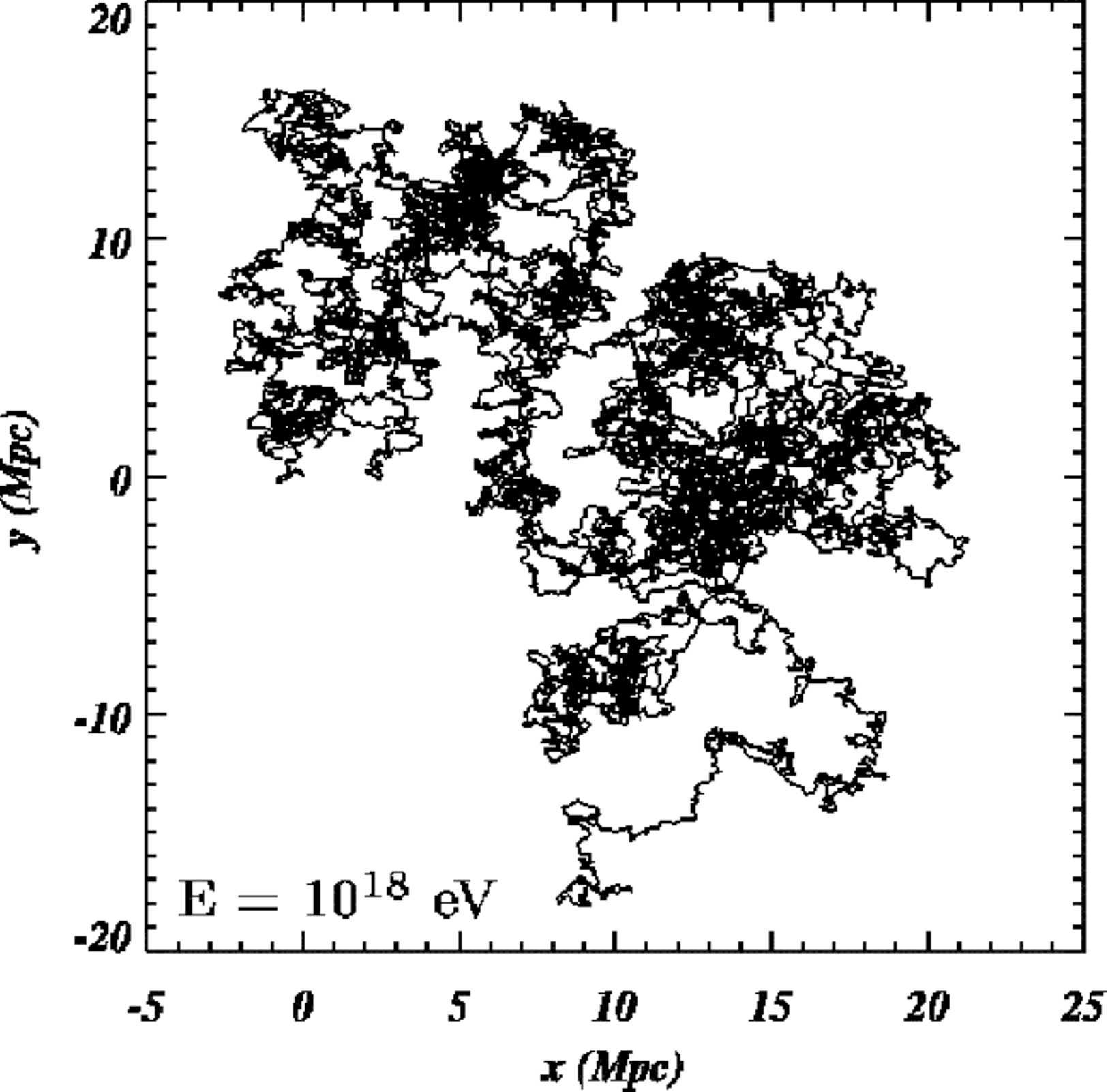}\hfill
\includegraphics[width=0.24\textwidth]{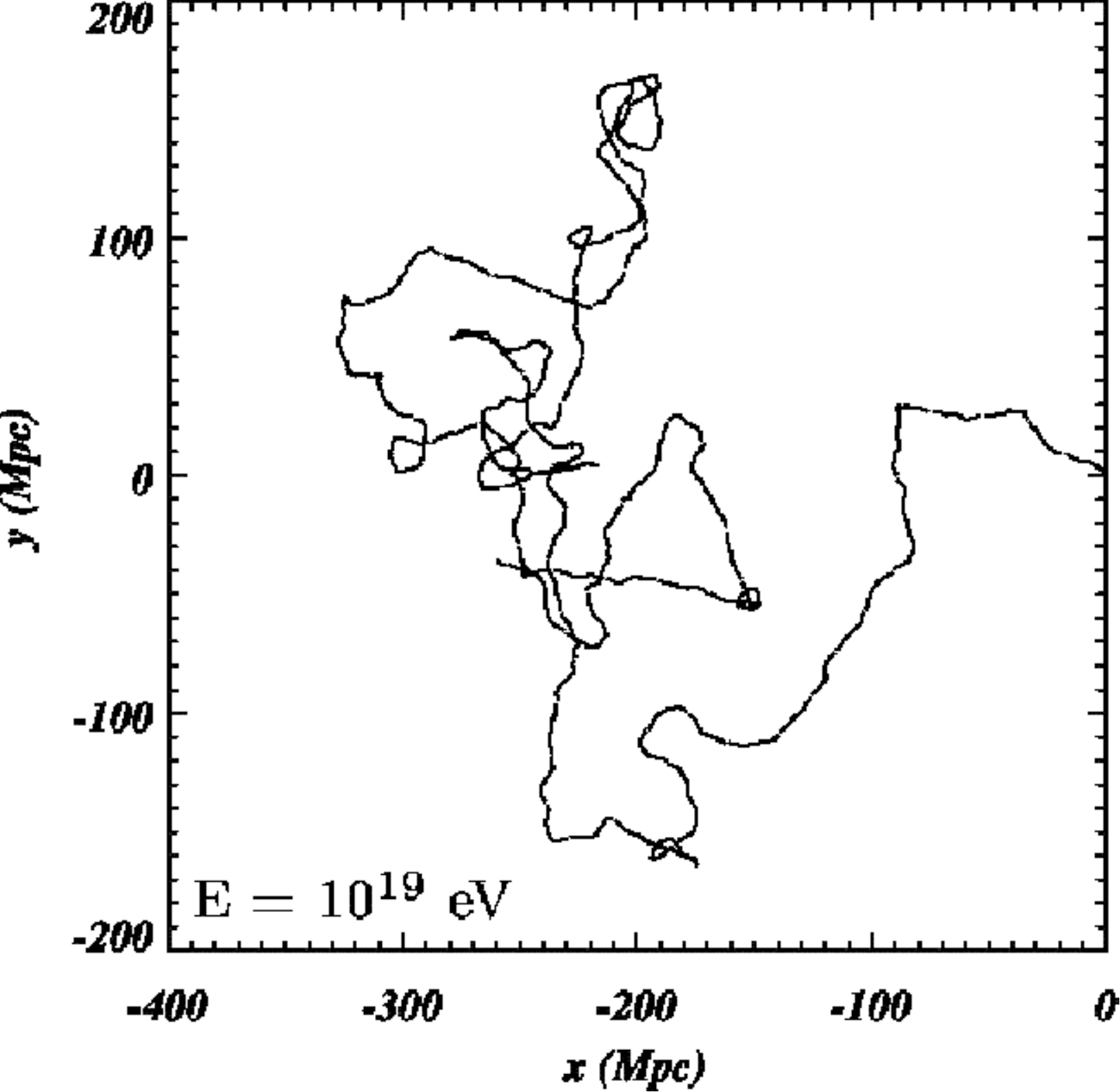}\hfill
\includegraphics[width=0.24\textwidth]{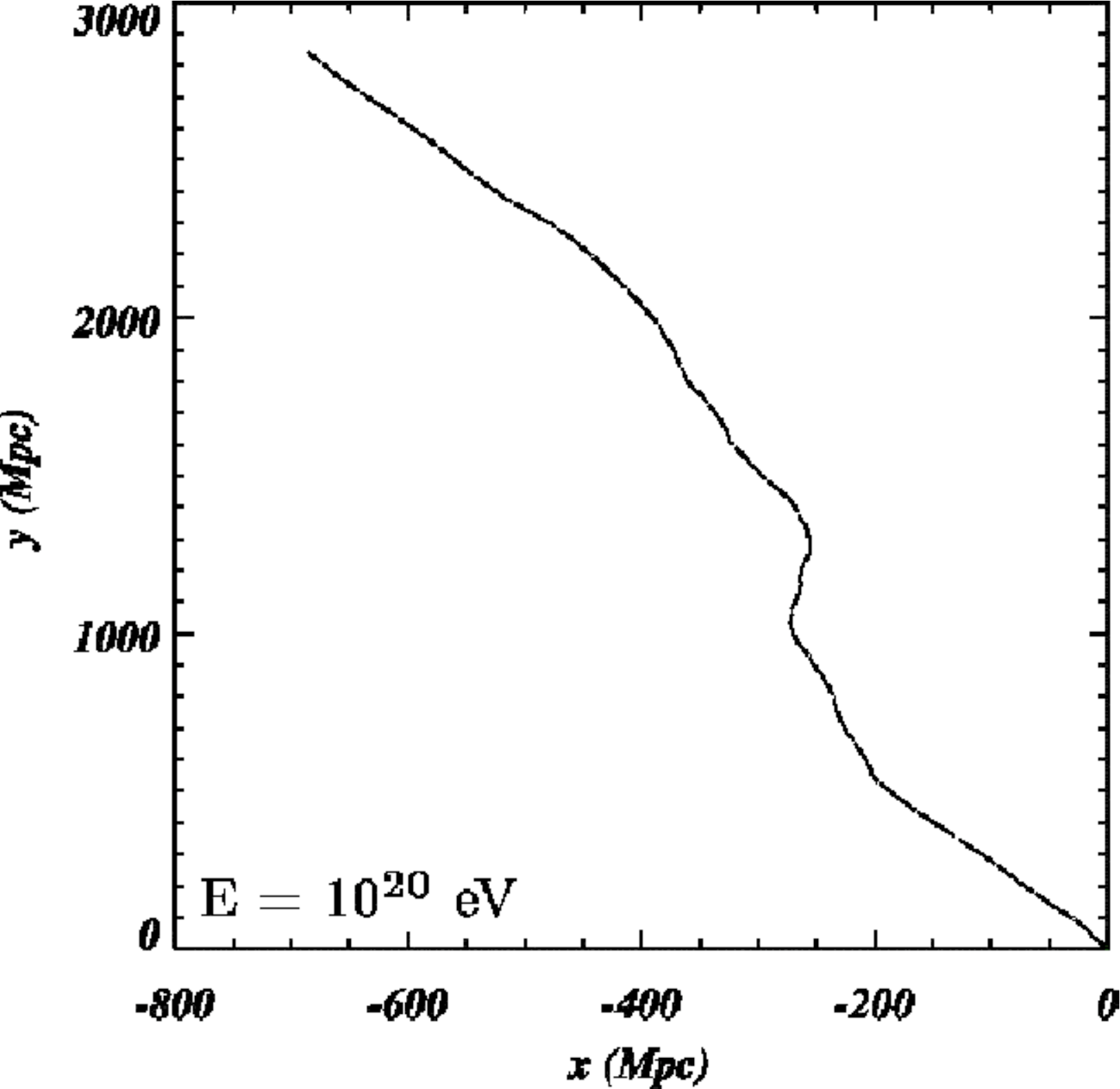}}
\caption{Examples of trajectories of protons of $10^{17}$~eV, $10^{18}$~eV, $10^{19}$~eV and $10^{20}$~eV (from left to right) in a turbulent magnetic field with $B_{\mathrm{rms}} = 10$~nG and $\lambda_{\mathrm{max}} = 1$~Mpc.}
\label{trajectoires}
\end{figure*}

Figure~\ref{fig:fieldStructure} shows some structures of a particular realization of the turbulent field modeled by Eq.~(\ref{eq:BGC}), assuming a Kolmogorov-like turbulence with principal scale $\lambda_{\mathrm{max}} = 1$~Mpc and a field variance of $\sqrt{\langle \|\mathbf{\vec{B}^2}\| \rangle}$ = 10 nG (hereafter, we refer to this quantity as the r.m.s. value, $B$, or simply as ``the magnetic field intensity''). Naturally, the values of an individual component of the field are distributed over a Gaussian of mean 0, and the coherence length $\lambda_{\mathrm{c}}\simeq\lambda_{\mathrm{max}}/5 \simeq 0.2$~Mpc, as expected for a Kolmogorov-like turbulence (Harari et al., 2002).

Once the turbulent field is modeled, the spatial transport of charged particles is simply followed by integrating the Lorentz equation, $\gamma\frac{d\mathbf{\vec{v}}}{dt}=\frac{q}{m}\mathbf{\vec{v}}\times\mathbf{\vec{B}}$, which we performed with either the Bulirsch-Stoer or the forth order Runge-Kutta methods (Press et al., 1993 and references therein) with identical results in all respects.

\section{Transport of charged particles in turbulent magnetic fields}
\label{sec:transport}

\subsection{Transition from ballistic to diffusive regime}

The transport of charged particles in a turbulent field is conceptually very simple, since, as mentioned above, it comes down to integrating trajectories influenced by the sole Lorentz force. However, its detailed treatment is limited by the nature of the problem, which is essentially chaotic (and of course by our ignorance of the real configuration of the fields). We are thus forced to keep to a statistical description of the trajectories. To characterize the particle transport and its different regimes, we have simulated 100,000 trajectories for protons between $10^{16}$~eV and $10^{20}$~eV (5,000 trajectories for each energy considered) in a 10~nG field ($\lambda_{\mathrm{max}}$ = 1 Mpc) (see Fig.~\ref{trajectoires}). The energy loss processes were switched off in these simulations.

\begin{figure}[ht]
\centering{\includegraphics[width=\columnwidth]{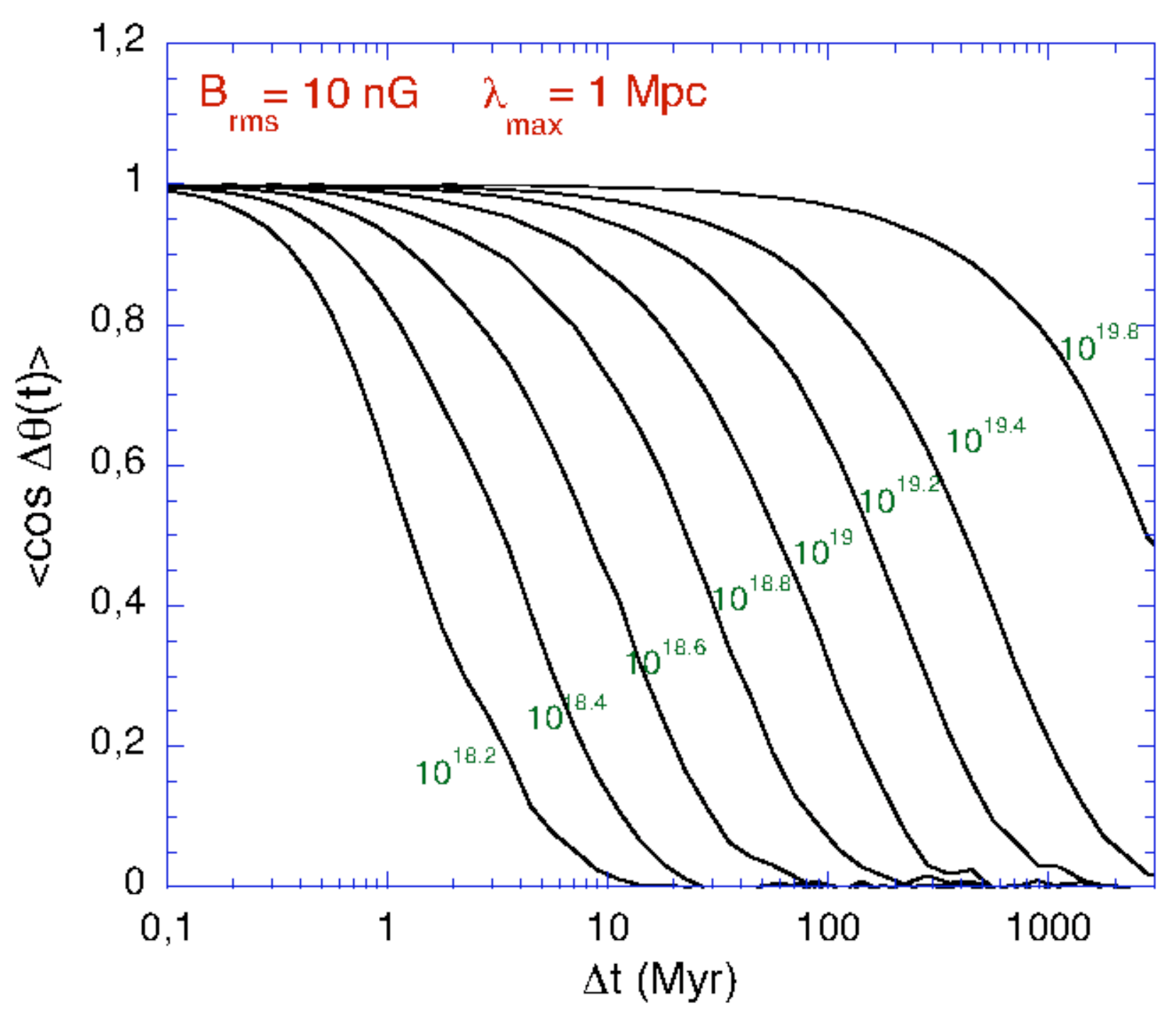}}
\caption{Angular correlation, $\left<\cos\delta\theta\right>$, of a set of protons propagating in a 10~nG turbulent field as a function of time, at different energies indicated by the labels (in eV).}
\label{Angular}
\end{figure}

One should first note that individual trajectories only depend on the particle gyroradius, $r_{\mathrm{g}}$, which can be expressed as a function of particle energy E, charge Z and magnetic field B, as:
\begin{equation}
r_{\mathrm{g}} \simeq 1.1\,\mathrm{Mpc} \times \frac{E_{\mathrm{EeV}}}{ZB_{\mathrm{nG}}}.
\label{eq:rG}
\end{equation}
In weak fields or at high energy (cf. Eq.~\ref{eq:rG}), when the particles have gyroradii much larger than the coherence length, $\lambda_{\mathrm{c}}$, they are only slightly deflected over that length, with $\delta\theta(\lambda_{\mathrm{c}})\simeq r_{g}/\lambda_{\mathrm{c}}$. Since the direction of the deflections are not correlated from one coherence length to the next, particles diffuse in angle (relative to the initial direction), with an angular spread $\sigma_{\theta}\sim t^{1/2}$.  It also appears that the \emph{angular decorrelation} (i.e, the loss of memory of the initial direction) occurs on a longer time-scale for higher energy particles, because of their larger rigidity. Figure~\ref{Angular} shows the evolution of the angular correlation $\zeta(t)=\mathbf{\vec{u_{\mathrm{0}}}}.\mathbf{\vec{u}}(\mathrm{t})$ (where $\mathbf{\vec{u_{\mathrm{0}}}}$ is the initial velocity vector and  $\mathbf{\vec{u}}(\mathrm{t})$ the velocity vector after a propagation time t) as a function of time, for different proton energies. One can see that the decorrelation process spreads over typically one order of magnitude in energy, and two orders of magnitude in time.

\begin{figure}[ht]
\centering{\includegraphics[width=\columnwidth]{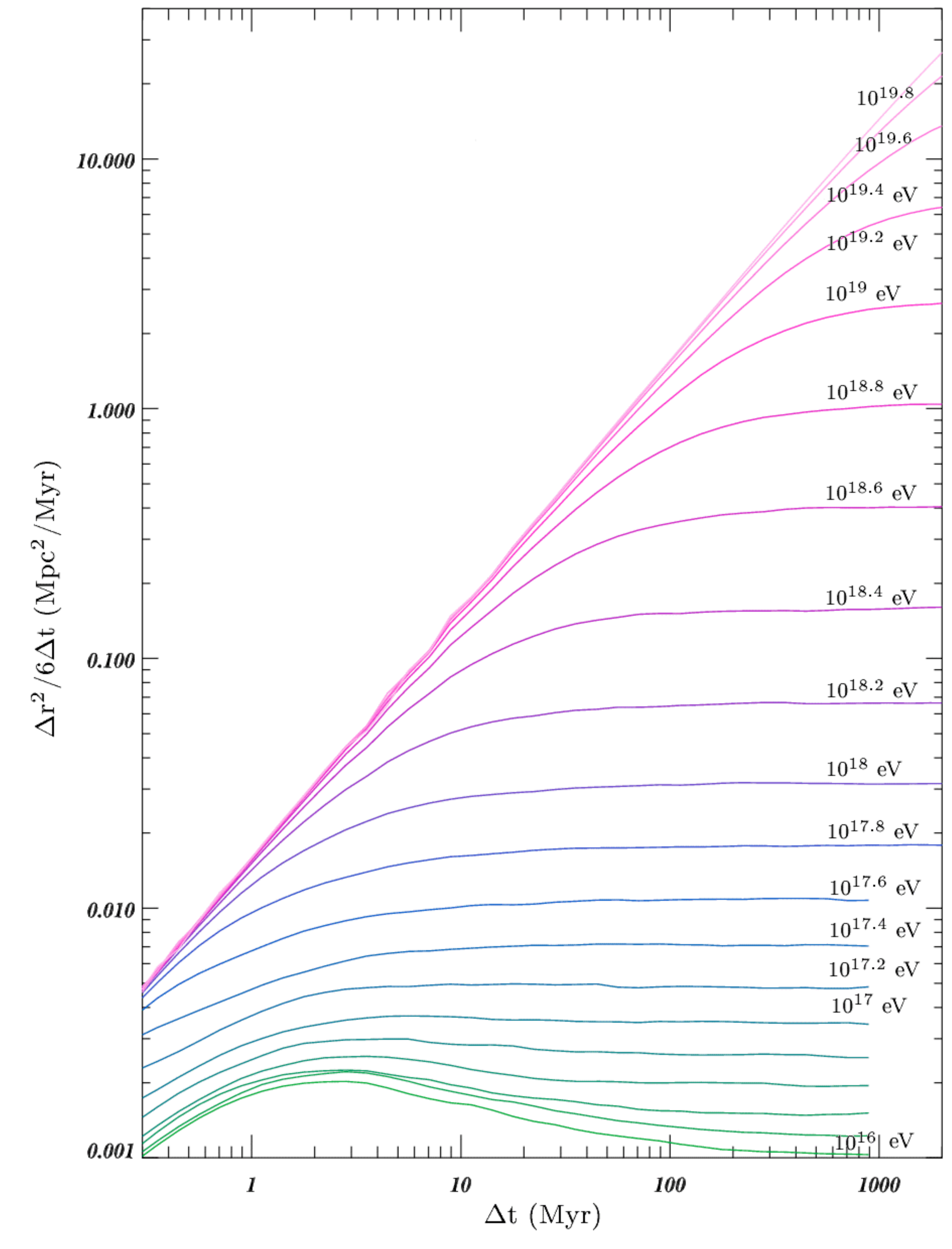}}
\caption{Instantaneous effective diffusion coefficient, $\Delta r^2/6\Delta t$, as a function of elapsed time, for protons of various energies, from $10^{16}$ to $10^{20}$~eV. The magnetic field is turbulent with an r.m.s. value of $B = 10$~nG and principal scale $\lambda_{\mathrm{max}}$ = 1 Mpc. Energy losses have been switched off.}
\label{fig:IECD}
\end{figure}
\begin{figure}[ht]
\centering{\includegraphics[width=\columnwidth]{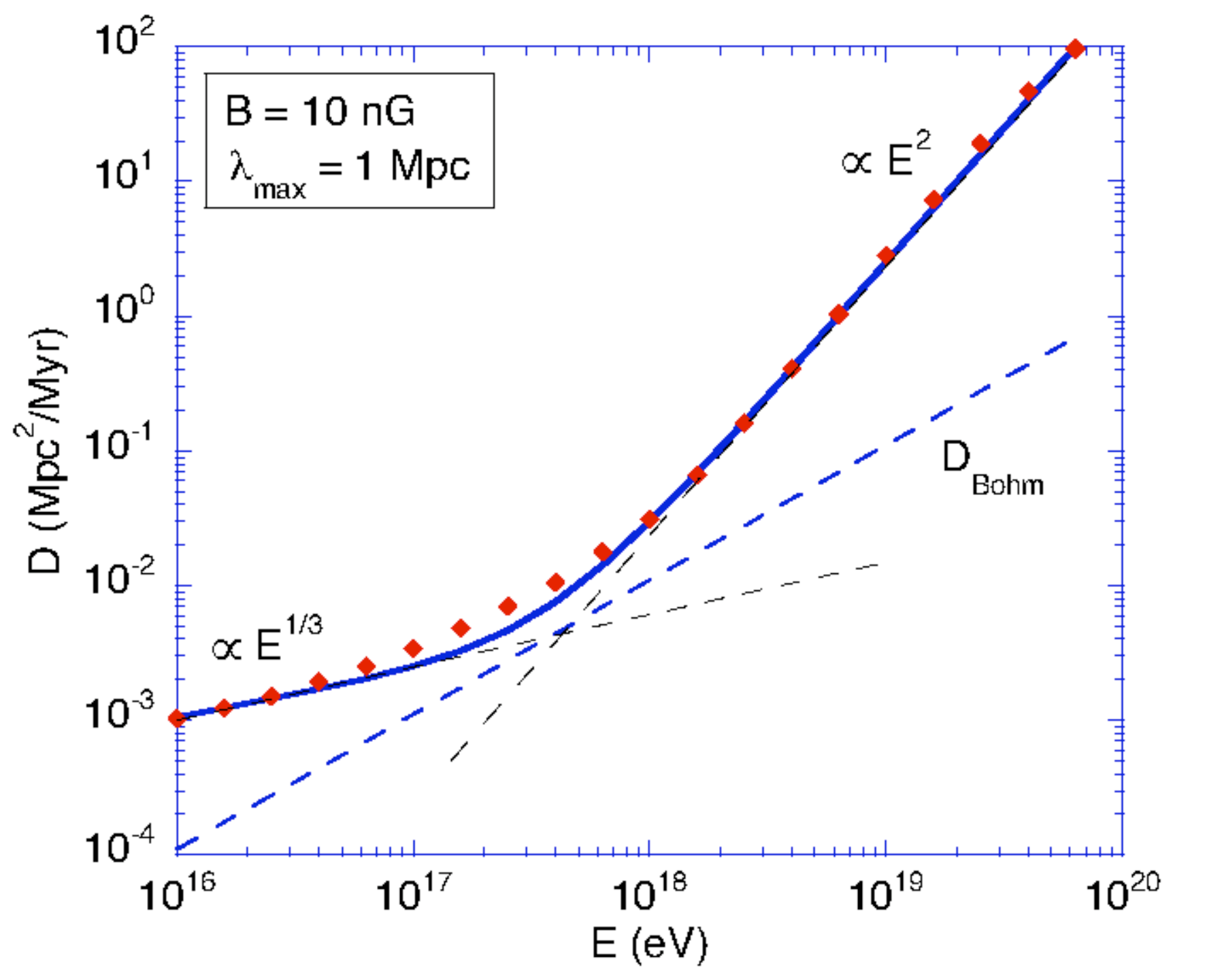}}
\caption{Diffusion coefficient of protons as a function of energy, for a turbulent margnetic field with r.m.s. value $B = 10$~nG and $\lambda_{\mathrm{max}} = 1$~Mpc. The red dots are the computation results, with asymptotic behaviors indicated by the thin dashed lines. The thick blue line is the simple fit proposed in the text. The Bohm diffusion coefficient is also shown for comparison (dashed line).}
\label{fig:diffusionCoefficient}
\end{figure}

Once the particles have lost the memory of their initial direction and are completely isotropized, it is appropriate to abandon the description in terms of individual trajectories, and consider their propagation as a diffusion process, in which the average linear distance traveled by CRs increases as $\Delta r^2 = 6D\Delta t$ (this equation actually \emph{defines} the diffusion coefficient, $D$). In the same spirit as in (Casse et al., 2002), we calculated the average quantity $\Delta r^2/6\Delta t$ -- which may be called the \emph{instantaneous effective diffusion coefficient} (IEDC) -- for the 5,000 trajectories at each energy and plot the result as a function of elapsed time, $\Delta t$. Results are shown on Fig.~\ref{fig:IECD}. The IEDC first increases linearly with time (as CRs propagate in straight lines with velocity $c$), and then a transition occurs to a diffusive regime where the IEDC is constant, with a value which can be identified with the diffusion coefficient, $D(E)$. Of course, the time at which the diffusion regime settles and the value of $D$ depend on the CR energy: lower energy particles enter the diffusion regime earlier and have a smaller diffusion coefficient. The diffusion regime can usually be modeled as a random walk at constant velocity (here $v = c$) with an isotropic redistribution of the direction after a ``mean free time'', which can be identified to the scattering time, $\tau_{\mathrm{s}}$, after which the direction of the particle is (on average) decorrelated from the initial direction (see Fig.~\ref{Angular}). Quantitatively, the diffusion coefficients are given by the values of the plateau in Fig.~\ref{fig:IECD}. They are gathered in Fig.~\ref{fig:diffusionCoefficient}, showing the smooth transition between the low-energy \emph{quasi-linear regime}, where $D(E)\propto E^{1/3}$, and the high-energy \emph{non-resonant regime}, where $D(E)\propto E^2$, in excellent agreement with theoretical expectations. The transition between the two regimes of diffusion spreads over an order of magnitude in energy, around an energy $E_{0}$ such that $2\pi r_{\mathrm{g}}(E_{0}) = \lambda_{\mathrm{c}}$. A simple fit of the diffusion coefficient is represented on Fig.~\ref{fig:diffusionCoefficient}, corresponding to the following formula: $D(E) = D_{\mathrm{Bohm}}(E_{0})(E/E_{0})^{1/3} + D_{\mathrm{Bohm}}(E_{1})(E/E_{1})^2$, where $E_{1}$ is such that $2\pi r_{\mathrm{g}}(E_{1}) = \frac{3}{2}\lambda_{\mathrm{c}}$. The Bohm diffusion coefficient, $D_{\mathrm{Bohm}}(E)=\frac{1}{3}r_{\mathrm{g}}(E)c$, is also shown for comparison.


As shown above, the propagation of CRs in a magnetic field smoothly passes from a ballistic regime, where the deflections are small and the distance traveled by a CR away from its source grows as $r\sim ct$, to a diffusive regime where it goes (on average) as $r\sim\sqrt{4Dt}$ (we arbitrarily use the numerical factor 4 for aesthetic reasons). In principle, the diffusion regime is always reached, provided one waits long enough (a few $D/c^2$). To be specific, one may define $\tau_{\mathrm{diff}}(E)\equiv 4D(E)/c^2$ as the time required to reach the diffusion regime. Likewise, $\lambda_{\mathrm{diff}}(E)\equiv c\tau_{\mathrm{diff}} = \sqrt{4D\tau_{\mathrm{diff}}} = 4D(E)/c$ is the distance traveled away from the source before CRs of energy $E$ isotropize and diffuse.

\begin{figure}[t]
\centering{\includegraphics[width=\columnwidth]{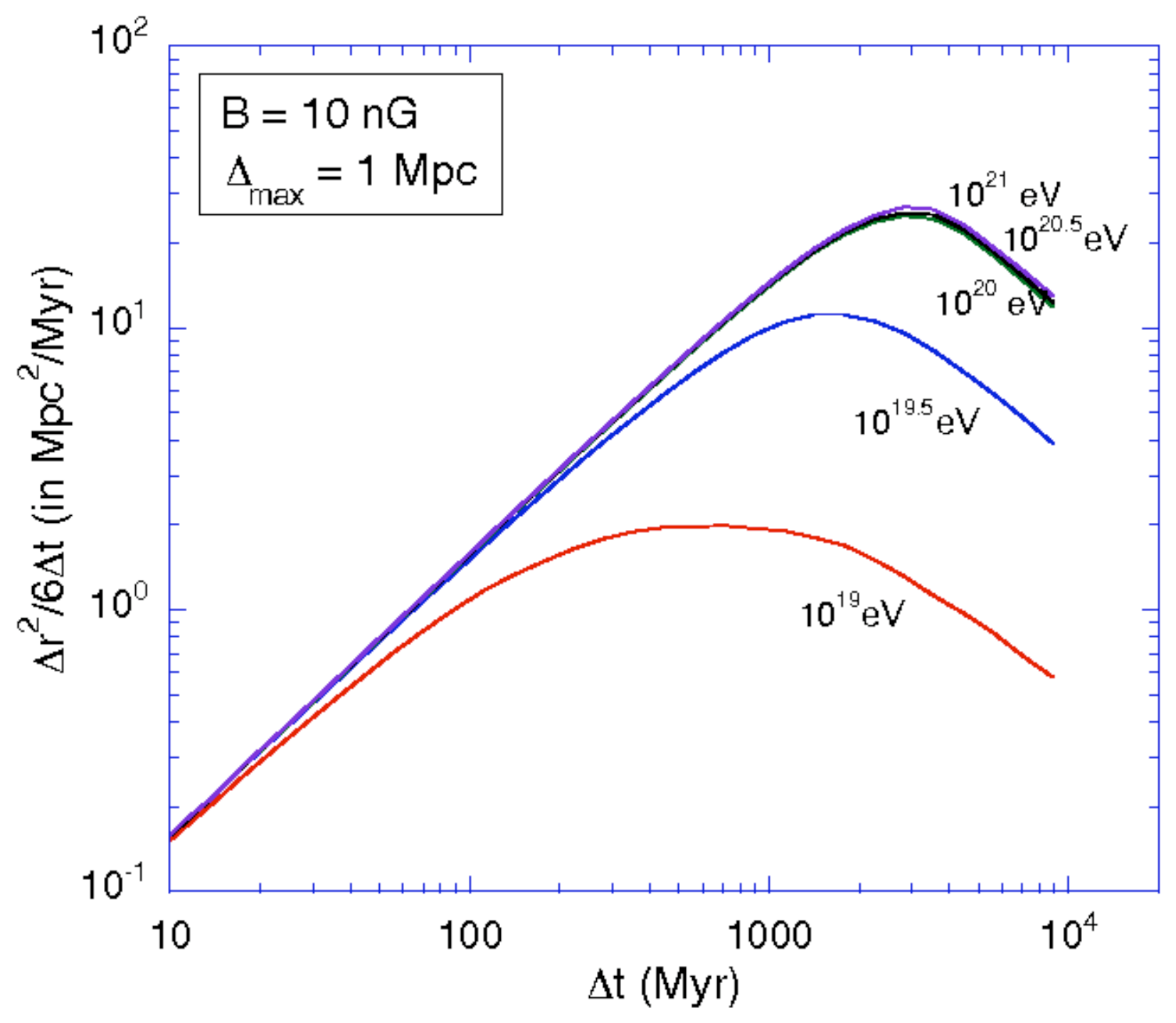}}
\caption{Same as Fig.~\ref{fig:IECD}, with energy losses switched on.}
\label{fig:IEDCWithLosses}
\end{figure}

However, CRs may not ``survive'' long enough to enter the diffusion regime. When proton energy losses are included in the propagation code, the IEDC curves look like in Fig.~\ref{fig:IEDCWithLosses}. A diffusion plateau is never reached, because the particles lose energy and their instantaneous diffusion coefficient then drops to lower and lower values. With the assumed field parameters, all particles above $10^{20}$~eV have the same transport properties, because they pass below the photo-pion production threshold while still in the ballistic regime.

More precisely, we define a distance, $D_{90}(E,A)$, to represent the effective ``energy-loss horizon'' associated with a particle that is \emph{observed} (at the Earth) at energy $E$ and mass $A$ (while it was emitted at $E_{0}$ with mass $A_{0}$ by its source). Specifically, $D_{90}(E,A)$ is the distance such that 90\% of the particles observed at $E$ and $A$ have travelled a (curvilinear) distance smaller than $D_{90}$ from their source. Obviously, this effective distance depends in principle on the energy spectrum and cosmic-ray composition at the source and on the redshift distribution of the sources contributing to the observed flux.

\begin{figure}[t]
\centering{\includegraphics[width=\columnwidth]{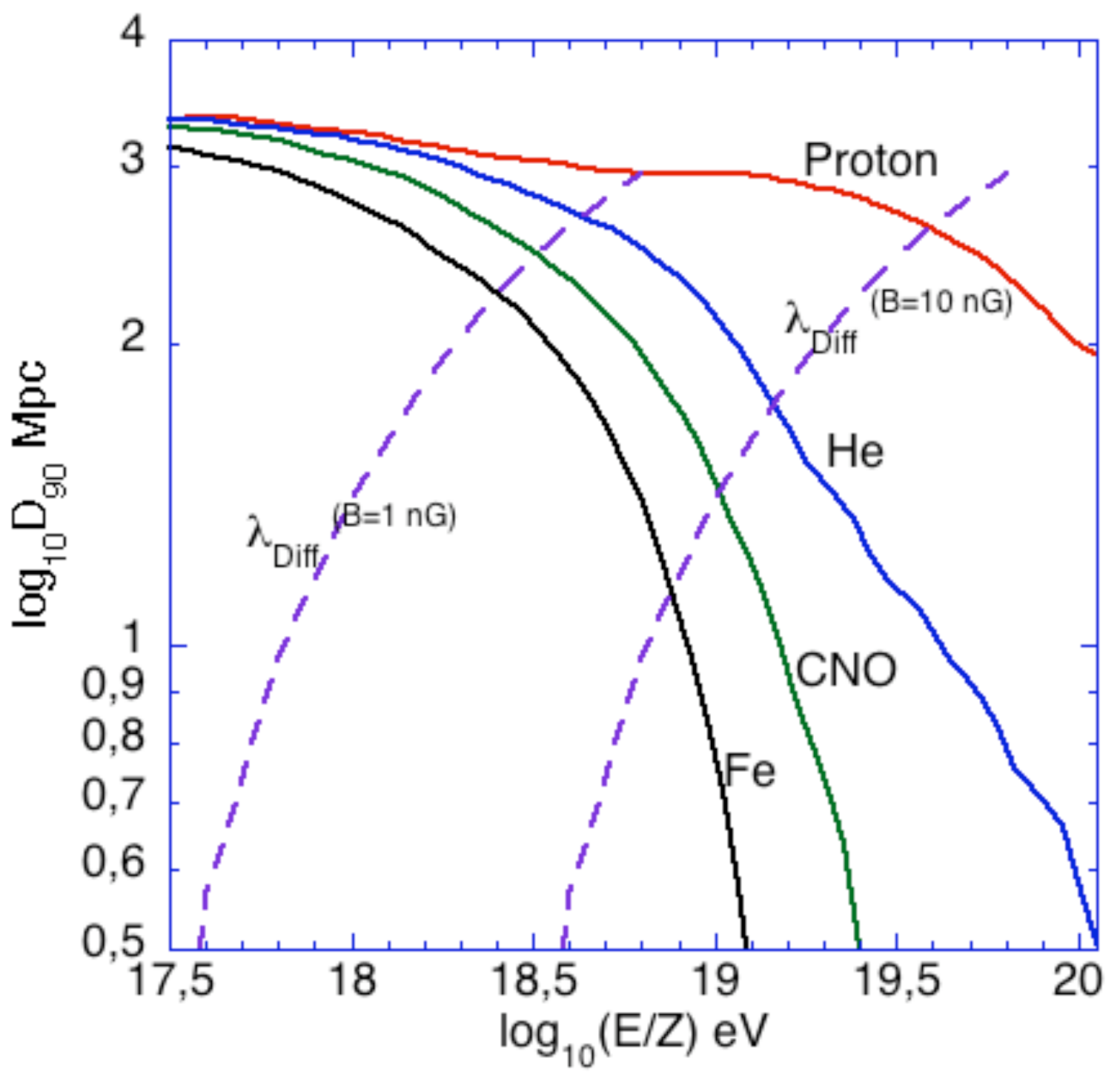}}
\caption{Effective energy loss horizon, $D_{\mathrm{90}}(E,A)$, for different nuclei as a function of $E/Z$, compared with the diffusion distance $\lambda_{\mathrm{diff}}(E/Z)$ for 1~nG and 10~nG magnetic fields.}
\label{fig:TdiffVsLoss}
\end{figure}

To estimate $D_{90}(E,A)$, we simulated the propagation of an extragalactic mixed composition with a spectrum in $E^{-2.4}$ (Allard et al., 2005), assuming a uniform source distribution. Figure~\ref{fig:TdiffVsLoss} shows the obtained $D_{90}$ as a function of $E/Z$ for protons, He, CNO and Fe nuclei, compared with $\lambda_{\mathrm{diff}}$ for magnetic fields of 1~nG and 10~nG. The diffusive approximation typically holds, for a given value of the magnetic field, whenever $\lambda_{\mathrm{diff}}\la D_{\mathrm{90}}(E,A)$. This depends not only on $B$ and $E/Z$ (as would be the case without energy losses), but also on the mass of the particle.

Interestingly, for a 10~nG field, the transition between ballistic and diffusive regimes for protons in the GZK range ($3\,10^{19}$--$10^{20}$~eV), occurs just around the \emph{GZK horizon}, as illustrated in Fig.~\ref{fig:TdiffVsLoss}. This change of propagation regime in the GZK region of the CR spectrum is one of the ingredients of a possible alteration of the standard GZK feature, as we now briefly discuss (see also Deligny et al., 2004).

\subsection{Magnetic horizons and modification of the GZK argument}

To see how magnetic fields can modify the appearance of the CR spectrum around the GZK feature, usually calculated with $B = 0$, it is useful to introduce the concept of \emph{magnetic horizon}, i.e. the maximum distance which (most) CRs can travel away from their source in a given magnetic field. This is set by the energy loss time horizon, estimated here as $\tau_{\mathrm{90}} = D_{\mathrm{90}}/c$ -- or the age of the source, $t_{\mathrm{s}}$, if it is smaller --, and the diffusion coefficient (or the IEDC if the diffusion regime is not reached). For high energy particles, roughly propagating in straight line, the magnetic field has no influence and one simply gets the usual GZK horizon. For low energy particles, the diffusion process prevents significant propagation beyond the magnetic horizon:
\begin{equation}
R_{\mathrm{magn}}(E) \simeq \sqrt{4D(E)~\mathrm{min}(\tau_{\mathrm{90}}(E),t_{\mathrm{s}})}.
\label{eq:RMagn}
\end{equation}
According to the standard GZK argument, the CR flux above $10^{20}$~eV should be sharply reduced compared to that below $3\,10^{19}$~eV, say, because of the sudden decrease of the GZK horizon distance. However, in the presence of relatively strong magnetic fields, low-energy particles \emph{also} have a limited range. In this sense, magnetic fields behave as ``low-cut filters'', whereas the CMB makes the universe a ``high-cut filter'' (GZK effect). The association and tuning of such filters can lead to various shapes of the propagated CR spectrum, different from the simple ``universal'' GZK feature.

In a 10~nG field, for instance, the magnetic horizon should severely limit the spatial propagation of low energy particles whereas high energy particles (above $5\,10^{19}$ eV) propagate in straight line. One can then imagine a situation where the \emph{same number of sources} could contribute at $10^{19}$~eV and $10^{20}$~eV! This does not mean, however, that there should not be any GZK feature in that case. Indeed, while magnetic fields prevent CRs from diffusing far from their sources, they also increase the CR density around each source, i.e. within the magnetic horizon. This is nothing but the magnetic confinement effect, very familiar for Galactic CRs. Since the number of particles remains the same, both effects exactly compensate, provided that the \emph{magnetic confinement spheres} (of radius $r_{\mathrm{H}}$) centered on the different sources are big enough to merge. If this is the case, the propagated CR spectrum with and without magnetic field are identical: the spatial horizon of the original GZK argument simply translates into a \emph{time horizon}, CRs at $10^{19}$~eV being able to contribute for a much longer time to the observed flux than CRs above $10^{20}$~eV. In fact, the propagated spectrum only depends on the time-of-flight distribution of the detected CRs, and it obviously makes no difference whether the trajectories were curved or not (see the formal demonstration of the ``propagation theorem'' in Aloisio and Berezinsky, 2004).

If the magnetic confinement spheres do \emph{not} fully merge, however, interesting spectral effects can appear. The exact shape of the UHECR spectrum thus depends crucially on the source granularity, i.e. the typical distance between the most important sources, $\Delta R_{\mathrm{s}}$ (see Deligny et al., 2004; Aloisio and Berezinsky, 2004). The attenuation of the expected GZK feature can then be expected for some particular combinations of the EGMF and source granularity. Let us note, however, that this scenario has been invoked to explain the absence of the GZK cut-off claimed by AGASA, but is now less attractive after the results of the HiRes (Thomson et al., 2006) and Auger (Yamamoto et al., 2007) experiments, finding a highly significant flux suppression above $\sim 5\,10^{19}$~eV.

\subsection{Effect of magnetic horizons at low energy}

Another effect related to the existence of magnetic horizons may arise  if we are outside the confinement sphere around a source, which is smaller and smaller as the energy decreases. In this case, particles emitted by that source simply could not reach us. The induced suppression of the flux at low energy could be of crucial importance (between $10^{17}$~eV and $10^{18}$ eV, or even above in the case of strong fields) and impact the transition between Galactic and extragalactic CRs. Such a scenario has been invoked in the context of second-knee transition model to lower the contribution of the extragalactic protons around $10^{17}$~eV. It has been shown (Lemoine, 2005; Aloisio et al., 2005) that, for a typical distance between sources of the order of 50~Mpc, a 1~nG field (with $\lambda_{\mathrm{max}}$=1 Mpc) could significantly suppress the extragalactic proton component below $10^{18}$~eV and thus avoid a too large proton fraction at $10^{17}$~eV.

In the next section, we study some potential effects of extragalactic magnetic fields on the cosmic-ray spectrum and composition around the ankle, in the case of an extragalactic mixed composition. The influence of the relevant physical parameters is examined.



\section{Calculation of the propagated EGCR spectrum and composition}

\subsection{Building a trajectory database}

Non-negligible magnetic fields make the computation of the extragalactic cosmic-ray spectrum conceptually and technically more complicated, because there is not anymore a one-to-one relation between the distance of a source and the time it takes a particle to reach the Earth after it is emitted. When magnetic fields are neglected, one can simply sum the contribution of all sources in the universe by integrating over the source distance, and for each source distance use the source power at the retarded time (i.e. the unique time of emission of the particles contributing to the observed flux), allowing for a possible evolution of the source, and of course apply the relevant energy losses, i.e. compute the injection energy of a particle now observed at energy $E$. If the extragalactic magnetic fields are not negligible, and especially when a diffusion regime can settle, high-energy particles emitted \emph{at a given cosmic time} -- i.e. redshift -- can be reaching the Earth \emph{now}, and thus contribute to the measured cosmic-ray flux, from sources in a large fraction of the universe.

Note that we make use of the word ``redshift'' here to represent a time, not a distance. Although these notions can be identified for a rectilinear propagation (photons or charged cosmic-rays with negligible magnetic fields), the most relevant notion for the present calculation is \emph{time}, since it parameterizes both the evolution of the sources and the changes in the conditions of propagation -- most notably the modification of the photon environment (CMB and infrared backgroud radiations) which control the energy losses of the cosmic rays. We will thus integrate over the time of emission of the particles, instead of the source distance.

The computation of the EGCR spectrum with magnetic fields can be formally simplified if one assumes that the diffusive regime of propagation holds for all particles at all energies. In that case, one can write down the diffusion equation (see Aloisio and Berezinsky, 2004 and Berezinsky and Gazizov, 2007, for a generalization to time-dependent energy losses and an expanding universe), and solve it  with the appropriate diffusion coefficients (see above). However, the highest energy particles cannot be expected to reach the diffusive regime before losing a significant fraction of their energy, even with magnetic fields near the highest values allowed by astrophysical observations. Using the diffusion equation at low energy and a standard rectilinear propagation at high energy would be appropriate in some cases, but the transition from one energy range to the other is likely to contain important information that would be missed with such an approach, and furthermore the transition energy should be different for different nuclear species. We thus chose to avoid approximations (either diffusive or rectilinear) and compute explicitly the trajectory of the particles in the EGMF, whatever their energy and mass, as described in Sect.~\ref{sec:transport}.

To save computational time, we developed a propagation scheme that effectively short-cuts most of the trajectory integration. The idea is to build a ``database'' for the particle transport, generated once and for all by integrating many trajectories with great accuracy over a limited time of propagation, and then simulate a realistic trajectory over a much longer timescale by putting one after the other the required number of (pre-computed) bits of trajectories, drawn randomly from the database. Of course, one can in fact forget everything about the trajectory itself, since we are only interested in the resulting energy, mass and position of a particle (with respect to its source) as well as its direction of propagation (with respect to its initial direction) after a given time of propagation. Joining random bits of trajectories one after the other, one obviously loses any correlation between them, so one needs to make sure that the bits of trajectory in the database are long enough for such correlations to be negligible. In practice, this is done by choosing trajectory lengths larger than the coherence length of the magnetic field, $\lambda_{\mathrm{c}}$ (see above). We checked that lengths of 1~Mpc and 10~Mpc gave the same results, and finally chose to divide the trajectories into steps of length $l_{\mathrm{step}} = 10$~Mpc (it takes longer to build the database than with 1~Mpc, but then the propagation runs to compute EGCR spectra and compositions are quicker). Note that we are referring here to curvilinear distances, i.e. the trajectories are integrated with full accuracy (using a Bulirsch-Stoer integrator to solve the equation of motion, with accuracies from $10^{-8}$ to $10^{-5}$) over a time $t_{\mathrm{step}} = l_{\mathrm{step}}/c \simeq 32.6$~Myr.

\begin{figure}[ht]
\centering{\includegraphics[height=6cm]{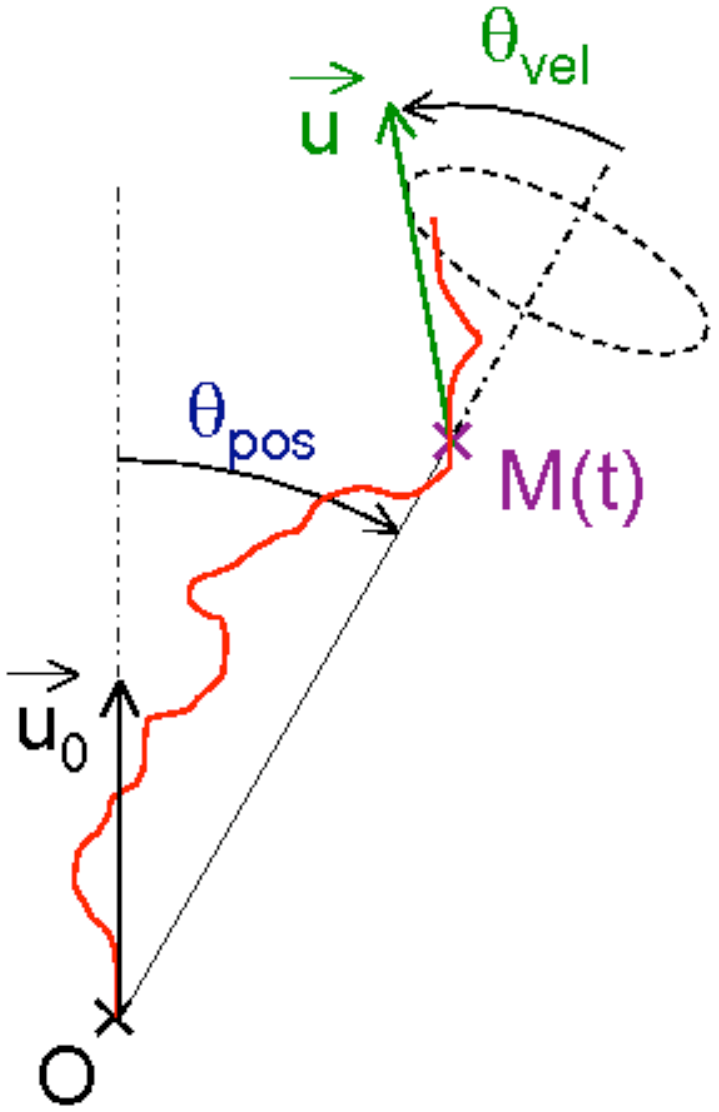}}
\caption{Geometry of the trajectories used in the fast integration method. The particle is initially injected in O with velocity $\mathbf{u}_{0}$. After a time $t$, it is in M$(t)$, at an angle $\theta_{\mathrm{pos}}$ from its original direction, and with a velocity $\mathbf{u}$ making an angle $\theta_{\mathrm{vel}}$ with the radial direction $\mathbf{OM}$ (azimuthal angles are chosen randomly assuming isotropy).}
\label{trajectoryGeometry}
\end{figure}

The database was built from $10^6$ proton trajectories, divided into 5,000 trajectories for each of the 200 initial energies considered, between $10^{16}$~eV and $10^{20}$~eV. The trajectories of nuclei with different charges, $Z$, are identical if the rigidities are the same, i.e. for identical $E/Z$ ratios. Each trajectory is summarized, after a propagation time $t$, by a triplet of quantities (see Fig.~\ref{trajectoryGeometry}): [$r(t),\cos\theta_{\mathrm{pos}}(t),\cos\theta_{\mathrm{vel}}(t)$], where $r(t) = \|\mathbf{OM}(t)\|$ is the \emph{rectilinear} distance traveled away from the source after a time $t$, $\theta_{\mathrm{pos}}$ is the global deflection with respect to the initial direction of propagation ($\mathbf{OM}\cdot\mathbf{u_{0}} = rc\cos\theta_{\mathrm{pos}}$), and $\theta_{\mathrm{vel}}$ indicates the current direction of propagation (with velocity $\mathbf{u}$) \emph{with respect to the radial vector} $\mathbf{OM}$ (not with the initial velocity). We verified that the corresponding azimuthal angles need not be registered for the purpose of our calculations, i.e. we can choose them randomly at each step (thus assuming isotropy in $\varphi_{\mathrm{pos}}$ and $\varphi_{\mathrm{vel}}$).

\begin{figure}[ht]
\centering{\includegraphics[width=\columnwidth]{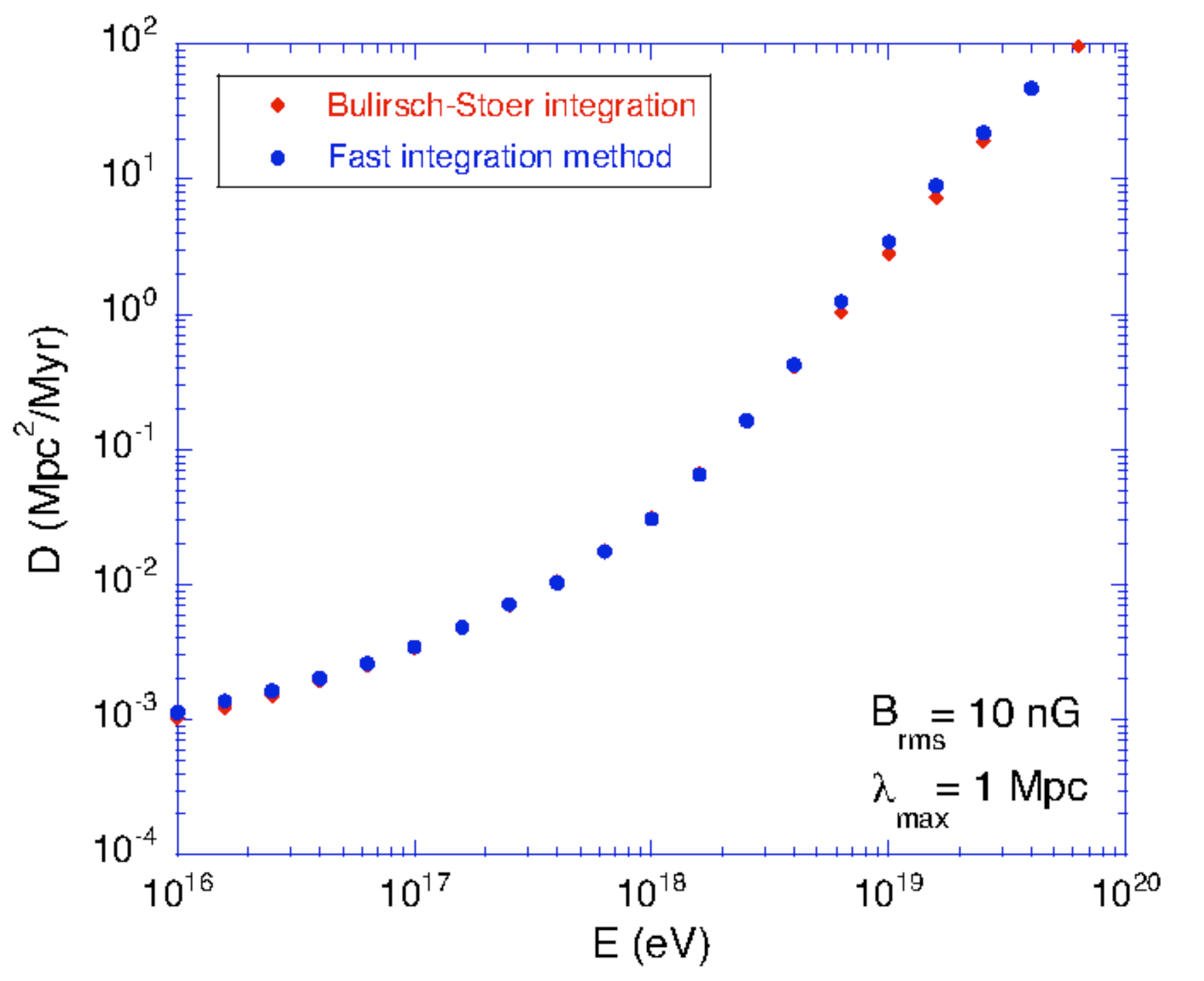}}
\caption{Comparison between the diffusion coefficients obtained with the detailed Bulirsch-Stoer integration (diamonds) and with our fast integration method (filled circles).}
\label{fig:DiffCoefComparison}
\end{figure}

The distribution of these (triplets of) quantities are tabulated for each class of trajectories (with a given rigidity), and used in the fast integration method as indicated above. An illustration of the efficiency of the method is given in Fig.~\ref{fig:DiffCoefComparison}, where we show the diffusion coefficients obtained with the fast method, compared with the original ones obtained with an accurate integration of the whole trajectories: the differences never exceed 15\%, which is negligible given our (absence of) knowledge of the magnetic fields exact structure and intensity.

Note that our main goal is to investigate and understand the potential effects of EGMFs on the high-energy cosmic-ray spectrum and composition, notably at the GCR/EGCR transition, not to work out a particular example based on specific assumptions about the structure of the magnetic field in our region of the universe. Therefore, we simply considered a stochastically homogeneous, turbulent magnetic field with r.m.s. intensities varying from 0 to 100~nG.

As a last step to a realistic simulation of the transport of high-energy particles in the extragalactic medium, we need to plug energy losses and nuclear photo-dissociation into the spatial trajectories. This is done using the interaction engine presented in (Allard, et al., 2006, 2007a, 2007b), by correcting the energy of the particle after each time step. This approximation was found to be convenient and reliable enough, as the potential change of energy \emph{during} a time step is negligible with our parameters. Indeed, the energy (or rigidity) loss time, $\tau_{\mathrm{loss}}$, of low-energy particles is much larger than the trajectory time step, while the highest energy particles with $\tau_{\mathrm{loss}}\gtrsim t_{\mathrm{step}}$ have essentially rectilinear trajectories and are thus not significantly affected by magnetic fields. As for nuclei, energy looses are mainly due to photo-disintegration processes, which do not alter their rigidity over a time $t_{\mathrm{step}}$. We could then check that the results of Fig.~\ref{fig:IEDCWithLosses} were reproduced accurately by our fast integration method.

\subsection{Propagated spectra from an individual source and a source distribution}

The method just described allows us to follow accurately a particle in energy space as well as in mass space and geometrical space (although only under the assumption of homogeneous, turbulent EGMFs), over a large period of time and eventually over the age of the universe or of the sources. Finally, to effectively compute the propagated EGCR spectrum and composition from a given source model (i.e. a given choice of the initial spectrum, initial composition, distribution of sources and source evolution), we need to identify which particles actually contribute to the cosmic rays observed \emph{now} and weight the effective energy and mass distribution contributed by each source according to its power at the time of emission.

As recalled above, contrary to the case of negligible magnetic fields, distant sources will contribute from a large period of time, not from the sole (rectilinear) retarded time. Each source is characterized by its injection spectrum and composition: the number of cosmic rays of nuclear species $i$ (with relative abundance $\gamma_{i}$ \emph{at a given rigidity} $E/Z_{i}$) emitted per second and per unit energy is given by:
\begin{equation}
\frac{\mathrm{d}^2N_{i}(E,t)}{\mathrm{d}t\mathrm{d}E} = Q_{i}(E,t) = \gamma_{i}Q(E/Z_{i},t) = \gamma_{i}Q(E/Z_{i})\times f(t),
\label{eq:CRInjection}
\end{equation}
where the injection spectrum, $Q(E/Z)$, is assumed to be constant in time and identical for all species (at fixed rigidity), and the dimensionless function $f(t)$ describes the evolution of the source power, usually parameterized by the redshift, $z$: $f(t) = g(z)$, where the relation between $z$ and $t$ is given by the underlying cosmological model (we assume a standard universe with $\Omega_{\mathrm{m}} = 0.3$, $\Omega_{\Lambda} = 0.7$ and $H_{0} = 71\,\mathrm{km\,s}^{-1}\mathrm{Mpc}^{-1}$). The spectrum $Q(E)$ is normalized to an injected power of 1~erg/s above a minimal energy $E_{\min}$:
\begin{equation}
\sum_{i} \int_{E_{\min}}^\infty Q_{i}(E)E\mathrm{d}E = 1\,\mathrm{erg\,s}^{-1},
\label{eq:injectionPower}
\end{equation}
so that the source evolution function $f(t)$ is directly related to the source luminosity at redshift $z(t)$, $\mathcal{L}_{\mathrm{s}}(z)$, through the simple relation $f(t) = \mathcal{L}_{\mathrm{s}}(z(t))/(1\,\mathrm{erg\,s}^{-1})$. Note that the nuclear abundances \emph{at a given energy} are related to the ``rigidity-abundances'' $\gamma_{i}$ through the actual shape of the injection spectrum: $\alpha_{i}\propto \gamma_{i}Q(E/Z_{i})/Q(E)$, i.e. $\alpha_{i}\propto\gamma_{i}Z_{i}^{x}$ if the EGCR source spectrum is a power-law, $Q(E) \propto E^{-x}$.

We now define a time-dependent ``contribution function'', $F(A,E,r;t)$, representing the contribution to the observed EGCR spectrum of the particles emitted by a source located at a distance $r$, at a given lookback time $t$ (corresponding to a redshift $z(t)$). To compute this function, we use a Monte-Carlo procedure, injecting isotropically a large number, $N_{0}$, of particles, distributed according to the assumed source spectrum and composition, and propagating them from the injection time $t$ to the present time (i.e. from redshift $z(t)$ to $z = 0$). After that time, the particles are spread over a large volume around the source, with a variety of masses and energies as a result of their interactions with the photon background. We then simply register their number density, energy spectrum and composition in different bins of distances from the source, $r_{i}$ (where isotropy is assumed, according to our simplifying assumptions). Specifically, we count the number of particles $N(A_{i},E_{j},r_{k};t)$ found with mass $A_{i}$ in the $j^{th}$ energy bin, i.e. between $E_{j}$ and $E_{j+1}=E_{j}+\Delta E_{j}$, and in the $k^{th}$ distance bin, i.e. between $r_{k}$ and $r_{k+1} = r_{k} + \Delta r_{k}$, from which we derive the \emph{contribution function}:
\begin{equation}
F(A_{i},E,r;t) = \frac{N(A_{i},E_{j},r_{k};t)}{\Delta E_{j}\times 4\pi r_{k}^2\Delta r_{k}\times N_{0}},
\label{eq:contributionFunction}
\end{equation}
for $E_{j}\leq E\leq E_{j}+\Delta E_{j}$ and $r_{k}\leq r\leq r_{k}+\Delta r_{k}$, where we divided by $N_{0}$ to obtain a differential density \emph{per injected particle}. Note that $F(A_{i},E,r;t)$ has the dimensions of an energy distribution function, $\mathrm{Eev}^{-1}\mathrm{cm}^{-3}$, and depends on the injection spectrum and composition.

To obtain a sufficiently precise description of this contribution function, we followed no less than $10^{10}$ trajectories (which would be impracticable without a fast integration scheme, as discussed above), with a variety of nuclei and initial energies ranging from $10^{17.5}$~eV to $10^{21}$~eV, and analyzed the particle distribution in mass and energy after propagation times ranging from 10~Myr to 13~Gyr. The procedure was repeated with different source spectra and r.m.s. values of the magnetic field.

Once the ``contribution function'' is known for a given model, the cosmic-ray flux contributed \emph{now} by an individual source at distance $R$ is obtained by a simple time integration over the whole period of activity of the source, weighting the contribution function at a given look-back time $t$ by the corresponding number of particles injected per second (since $F$ is normalized to one injected particle: see Eq.~\ref{eq:contributionFunction}). This is equal to the source power, $\mathcal{L}_{\mathrm{s}}(t)$, divided by the average energy of a particle at the source $\langle E\rangle_{\mathrm{s}} = \int_{E_{\min}}^{\infty}\sum_{i}Q_{i}(E)E\mathrm{d}E/\int_{E_{\min}}^{\infty}\sum_{i}Q_{i}(E)\mathrm{d}E$. Finally, the contribution of an individual source is given by:
\begin{equation}
\Phi(A,E) = \frac{c}{4\pi}\int_{0}^{t_{\max}} F(A,E,R;t)\times\frac{\mathcal{L}_{\mathrm{s}}(t)}{\langle E\rangle_{\mathrm{s}}}\times\frac{\mathrm{d}t}{1+z(t)}
\label{eq:sourceContribution}
\end{equation}
where the factor $c/4\pi$ simply converts a CR density into a flux (in $\mathrm{cm}^{-2}\mathrm{s}^{-1}\mathrm{sr}^{-1}\mathrm{EeV}^{-1}$), and the factor $1+z(t)$ takes into account the time dilation between the injection time and the observation time, associated with the universal expansion.

Finally, the total EGCR spectrum is obtained by summing the contributions of all the sources in the universe. In practice, we assume a given source density (number of ``standard candle'' sources per Mpc$^3$), draw randomly of set of source distances accordingly, and add the contribution of each source as given by Eq.~(\ref{eq:sourceContribution}) with the appropriate distance. Note that changing the source evolution model does not require to compute new ``contribution functions'', since any evolution function $\mathcal{L}_{\mathrm{s}}(z)$ can be implemented in the last (quick) integration step, Eq.~(\ref{eq:sourceContribution}).

\begin{figure}[t]
\centering{\includegraphics[width=\columnwidth]{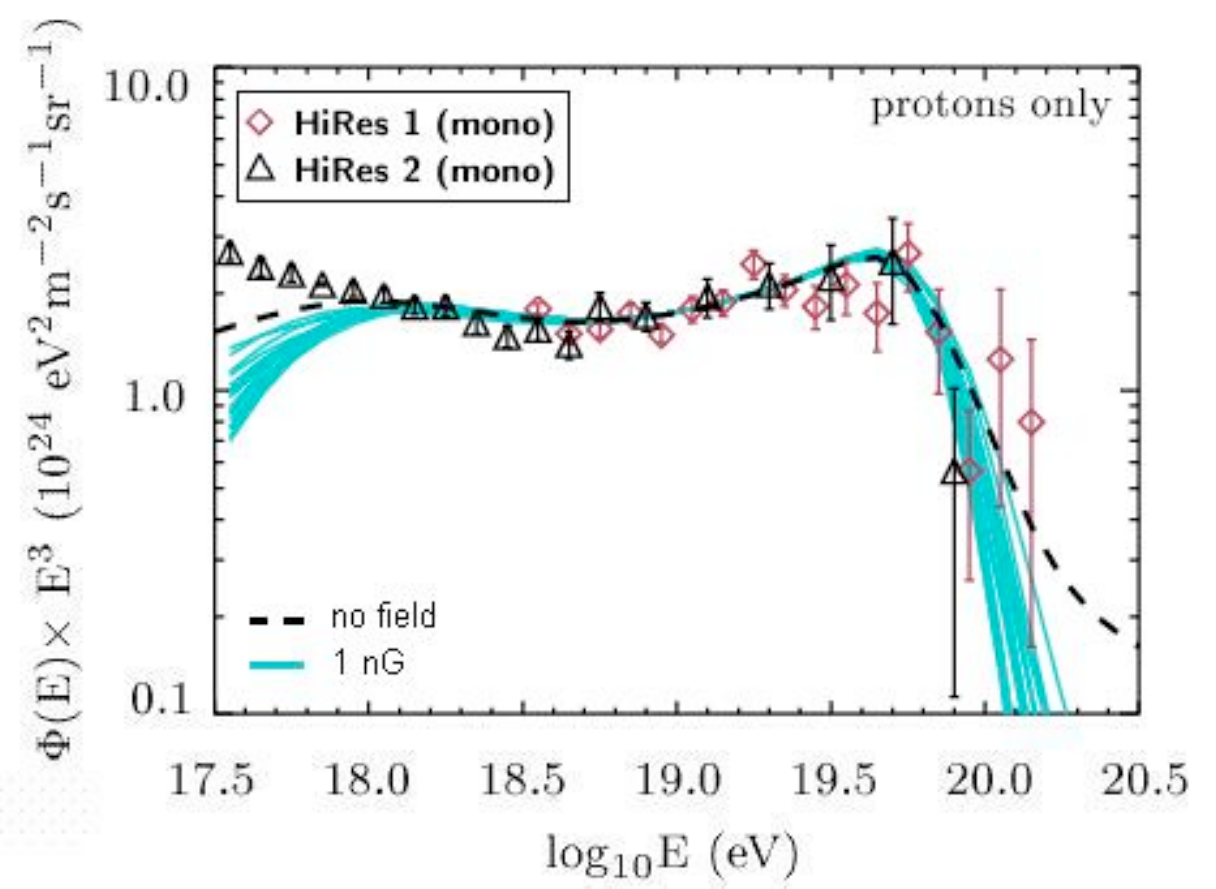}}
\caption{Propagated EGCR spectra for pure proton sources with an injection spectrum in $E^{-2.6}$, compared with the HiRes data (Bergman et al., 2005), for different realizations of the source positions, randomly drawn with a density of $n_{\mathrm{s}} = 10^{-5}\,\mathrm{Mpc}^{-3}$, and a magnetic field of $B = 1$~nG. The case of a continuous source distribution with $B = 0$ is also shown (dashed line).}
\label{fig:pureProtonSpectra}
\end{figure}

\section{Results}

\subsection{Pure proton sources}

Figure~\ref{fig:pureProtonSpectra} shows the propagated EGCR spectrum computed with the above method in the case of pure proton sources with a density of $10^{-5}\,\mathrm{Mpc}^{-3}$, injecting cosmic-rays with an initial power-law spectrum in $E^{-2.6}$ in a homogeneous turbulent extragalactic magnetic field with $B_{\mathrm{rms}} = 1$~nG  and $\lambda_{\max} = 1$~Mpc. Twenty different realizations of the local source distribution were considered, in order to explore the so-called \emph{cosmic variance}. The propagated spectra are normalized to fit the UHECR data from the HiRes experiment (Bergman et al., 2005). We also show the results obtained with negligible magnetic fields and a totally uniform source distribution (dashed line). As can be seen, the extragalactic spectrum is essentially unaffected above $10^{18}$~eV, apart from the sharper cut-off at high energy, due to the finite distance of the closest source in a realistic source model. At lower energy, the expected cosmic-ray flux is reduced as a result of the magnetic field horizon effect, and just as for the GZK cut-off above $10^{19.7}$~eV, the exact spectrum modification is subject to cosmic variance fluctuations, up to a factor of 2 in flux. These results are in good agreement with previous studies (Lemoine, 2005; Aloisio and Berezinsky, 2005; Berezinsky and Gazizov, 2007; Kotera and Lemoine, 2007).

In the following, we focus on the case when EGCR sources inject not only protons, but a mixed composition of nuclei.

\begin{figure}[t]
\centering{\includegraphics[width=\columnwidth]{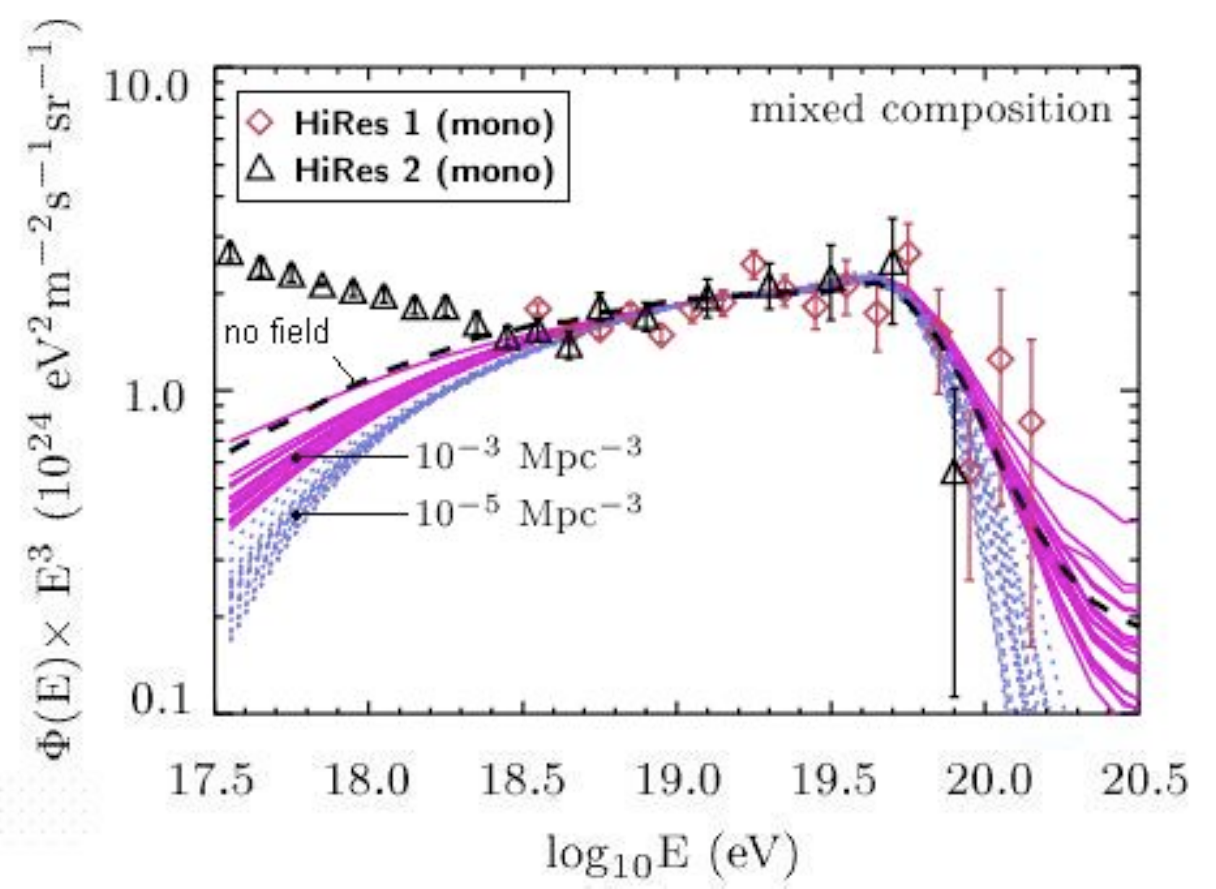}}
\caption{Propagated EGCR spectra for a mixed source composition and an injection spectrum in $E^{-2.4}$, compared with the HiRes data (Bergman et al., 2005), for different source densities and EGMF intensities: continuous source distribution (with $B=0$); $n_{\mathrm{s}} = 10^{-3}$ Mpc$^{-3}$ and $B = 1$~nG ($\lambda_{\mathrm{max}}$ = 1 Mpc);  $n_{\mathrm{s}} = 10^{-5}$ Mpc$^{-3}$ and $B = 1$~nG. Each line of a same group corresponds to a different Monte-Carlo realization of the source distribution.}
\label{fig:mixedCompositionSpectra_density}
\end{figure}

\subsection{Mixed composition sources}

As we have shown in previous works (Allard et al., 2005, 2007a, 2007b), the overall high-energy cosmic ray spectrum can be modeled as a superposition of contributions from sources distributed throughout the universe, injecting cosmic rays with a power-law spectrum in $E^{-x}$ and a mixed composition, similar to what is known from low-energy cosmic rays in our Galaxy. The required spectral index depends somewhat on the assumed source evolution function. A good fit of the data is obtained with a spectral index of $x \simeq 2.2$--2.3 if one assumes that the source power scales with the star formation rate as a function of time. Here, we focus on the influence of the magnetic field on the propagated EGCR spectrum, and simply assume that the source power remains constant, which favors a spectral index of $x \simeq 2.3$--2.4 (we choose $x = 2.4$, which gives better fits in the presence of a magnetic field).

\subsubsection{Magnetic horizons and EGCR contribution at the ankle}
\label{sec:magnHorizons}

Just as in the case of pure proton sources, and for the same reasons, the propagated mixed composition spectrum depends on the typical source density and on the particular distances of the most nearby sources (cosmic variance). This is shown on Fig.~\ref{fig:mixedCompositionSpectra_density}, where the EGCR spectra were computed assuming an EGMF of $B = 1$~nG and source densities of $10^{-3}\,\mathrm{Mpc}^{-3}$ or $10^{-5}\,\mathrm{Mpc}^{-3}$. The (unrealistic) case of a perfectly uniform source distribution is also shown for comparison (in that case, we set $B = 0$, since the magnetic field is irrelevant). As already noted in previous studies (Aloisio and Berezinsky, 2004; Berezinsky and Gazizov, 2007), it is natural that the propagated spectra obtained with increasing source densities converge towards the ``universal'' spectrum obtained assuming a continuous source distribution. For discrete sources, a flux deficit is expected at low energy as a result of the energy dependent magnetic horizon (see above), and for a given magnetic field intensity, its amplitude is larger for lower source densities (see Lemoine, 2005; Aloisio and Berezinsky, 2005; Kotera and Lemoine, 2007 for the study assuming the pure proton case).

\begin{figure}[t!]
\centering{\includegraphics[width=\columnwidth]{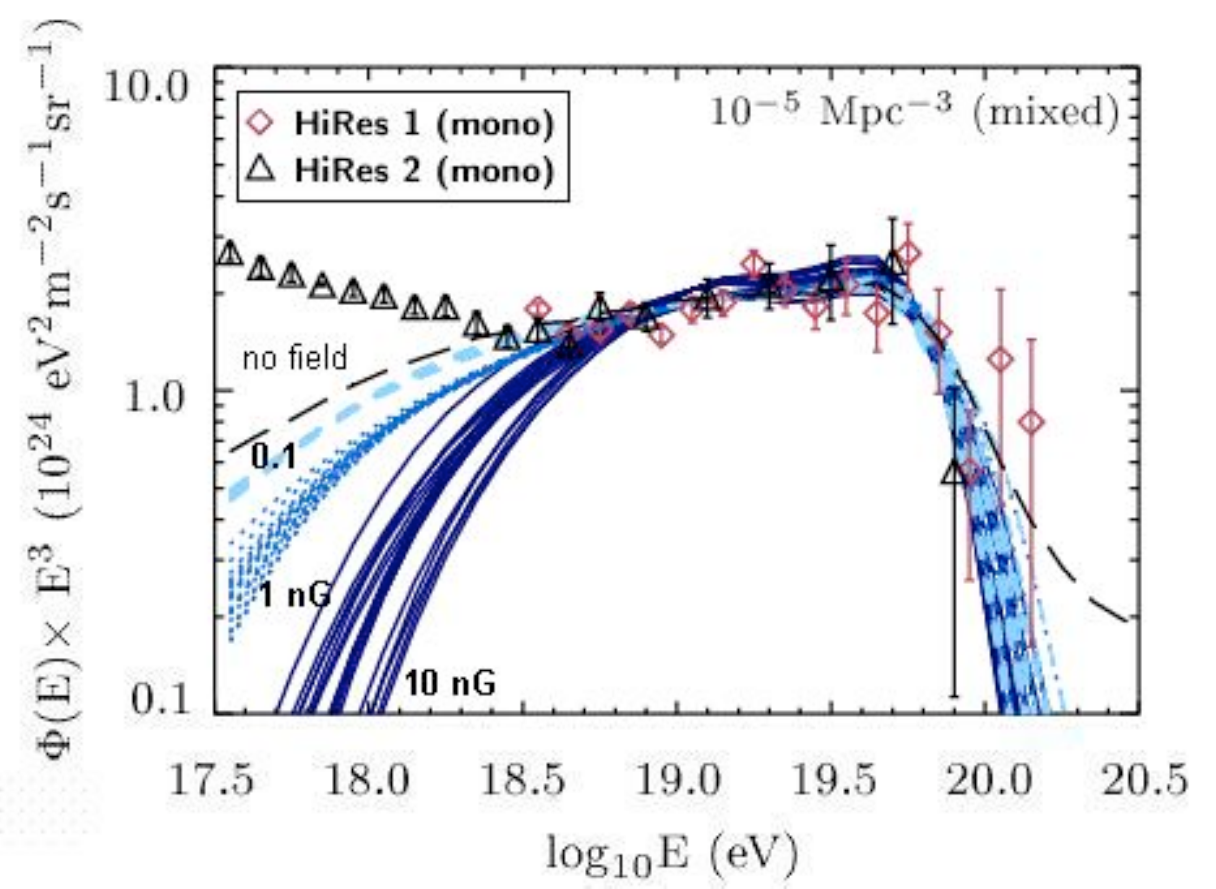}}
\caption{Propagated EGCR spectra for a mixed source composition and different magnetic field intensities: 0.1 nG ($x = 2.4$), 1 nG ($x = 2.4$) and 10 nG ($x = 2.6$ or 2.7, depending on which spectral index better fits the high-energy data, for the particular realization of the source distribution under consideration). In each case, $\lambda_{\mathrm{max}} = 1$~Mpc and $n_{\mathrm{s}} = 10^{-5}\,\mathrm{Mpc}^{-3}$.}
\label{spectre_field}
\end{figure}

On Fig.~\ref{spectre_field}, we also show the propagated spectra obtained for different Monte-Carlo realizations of a source density of $n_{\mathrm{s}} = 10^{-5}\,\mathrm{Mpc}^{-3}$, for different values of the magnetic field intensity: $B = 0$, $B = 0.1$~nG, $B = 1$~nG and $B = 10$~nG. Naturally, the magnetic horizon effect (depleting the cosmic-ray flux at low energy) is larger for larger field intensities. In the case of a 10~nG field (close to the upper limit determined by Blasi et al., 1999, although for a larger value of the field coherence length) the EGCR spectrum can be modified up to energies as high as $\sim10^{19}$~eV for the proton component (and of course more for heavier nuclei). Softer source spectra are thus needed to keep a good agreement with the data above the ankle, with power-law indices around 2.6--2.7 (the HiRes data are shown as an illustration). In any case, intense magnetic fields keep the contribution of the extragalactic component almost negligible up to $10^{18}$~eV and its composition very light above the ankle, since the heavy nuclei coming from even the closest sources cannot reach the Earth before being photo-disintegrated. Note that this combination of values of the source density and the magnetic field intensity implies a late transition from GCR to EGCR, i.e, a composition dominated by heavy galactic cosmic rays at least up to the ankle, which is compatible with classical ankle scenarios (see for instance Wibig and Wolfendale, 2004), but not favored by current composition analyses (see Sect.~\ref{sec:CRCompo}).

\begin{figure}[t]
\centering{\includegraphics[width=\columnwidth]{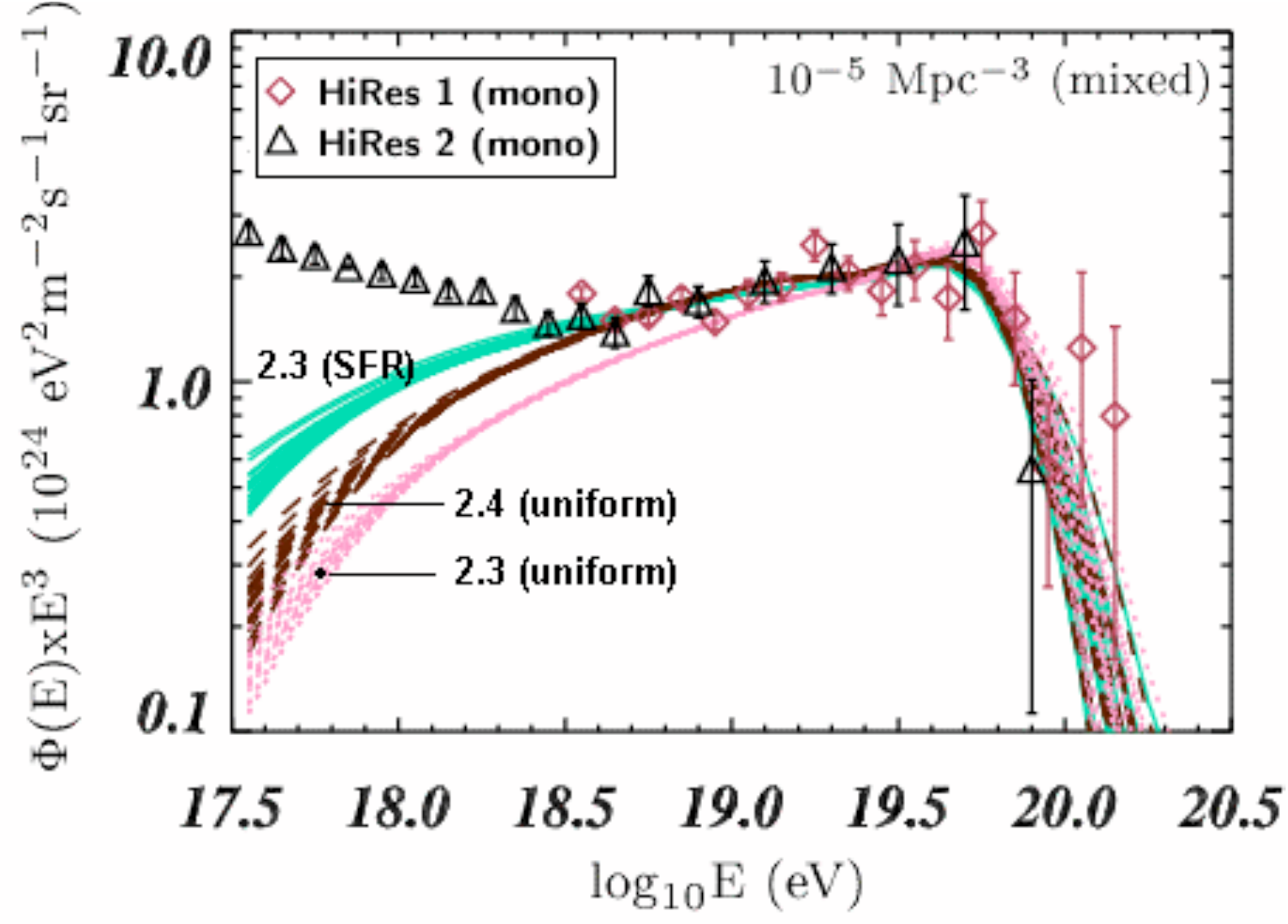}}
\caption{Propagated EGCR spectra for mixed composition sources with density $n_{\mathrm{s}} = 10^{-5}\,\mathrm{Mpc}^{-3}$, for various spectral indices and source power evolutions (the magnetic field intensity is $B = 1$~nG). From top to bottom:  $x = 2.3$ and a source power proportional to the star formation rate (SFR); $x = 2.4$ with no evolution; $x = 2.3$ with no evolution.}
\label{spectre_evol}
\end{figure}

The above results were obtained assuming sources without evolution (i.e with a constant luminosity and comoving density). Now, a larger source power in the past -- as expected, for instance, if the source density or intensity roughly scales with the star formation rate in the galaxies -- results in a larger contribution of the sources at lower energy, since the higher energy cosmic rays detected now were emitted more recently, from less intense sources. Therefore, in general, a harder source spectrum (i.e. with less contribution at low energy) is required to fit the high-energy CR data if source evolution is taken into account. Typically, a source spectral index of 2.2--2.3 instead of 2.3--2.4 is found to better reproduce the observed spectrum if the source power is proportional to the star formation rate (Allard et al., 2007b). 

In a magnetized universe, the source evolution can somehow compensate for the magnetic horizon effect. In Fig.~\ref{spectre_evol}, we show a few EGCR spectra propagated from mixed composition sources with a spectral index of $x = 2.3$, both without source evolution and with a source power proportional to the star formation rate (SFR), as estimated from Bunker et al.~(2004). Although different at low energy (where the EGCR component is subdominant), the latter case and the case of a uniform source distribution (i.e. without evolution) with a steeper spectral index of $x = 2.4$ both give acceptable fits of the HiRes data. We finally note that increasing the source power or the source density (number of sources per comoving volume) does not produce the same effect if the magnetic field is non-negligible, since a low source density may result in an almost complete absence of low-energy EGCRs at Earth, whatever the source luminosity, if the closest source lies outside the magnetic horizon at these energies. In this respect, an evolution of the source number density could in principle counterbalance the magnetic field suppression effect more efficiently than the evolution of the source luminosity.

\begin{figure*}[t]
\includegraphics[width=0.33\textwidth]{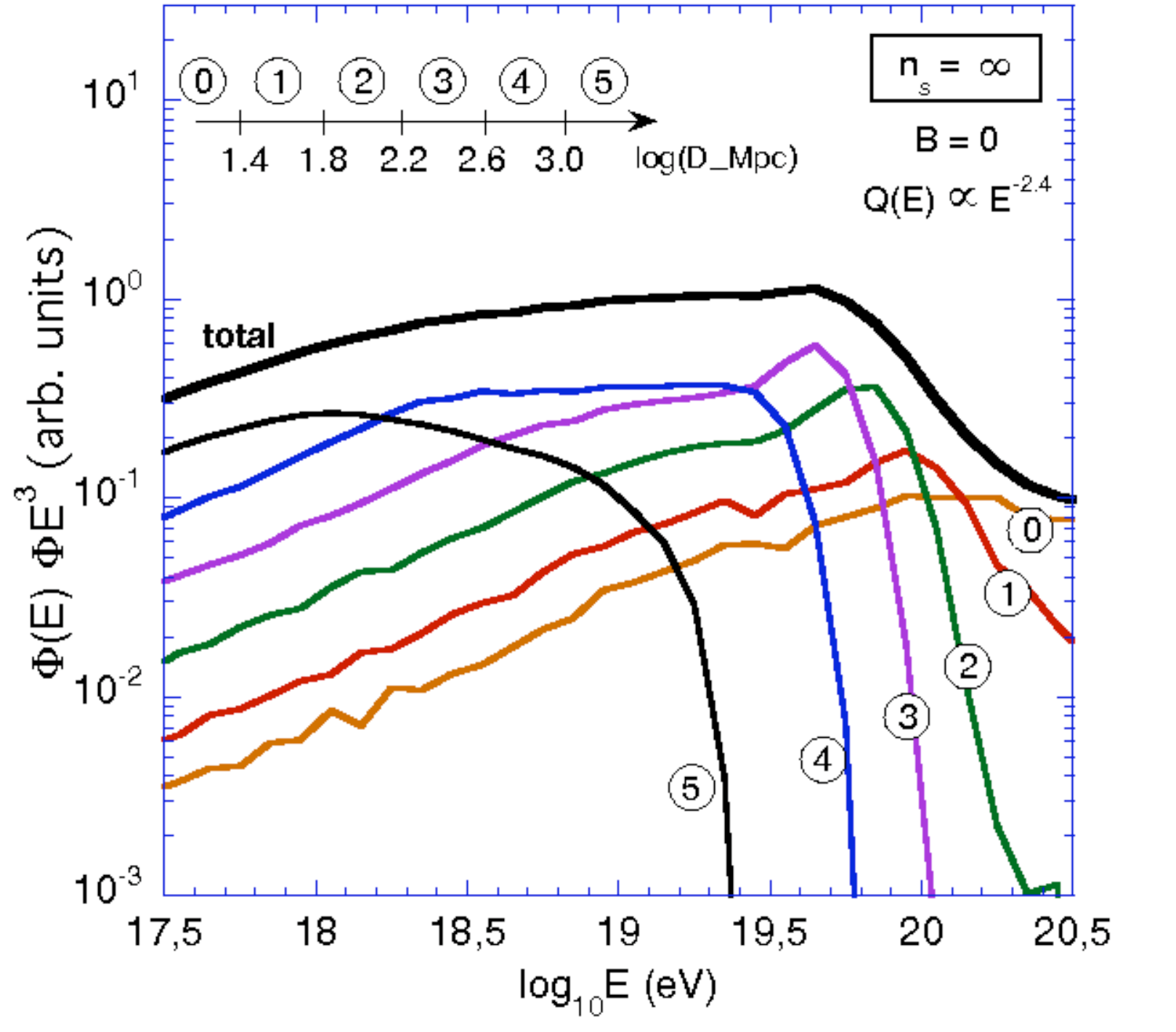}
\hfill \includegraphics[width=0.33\textwidth]{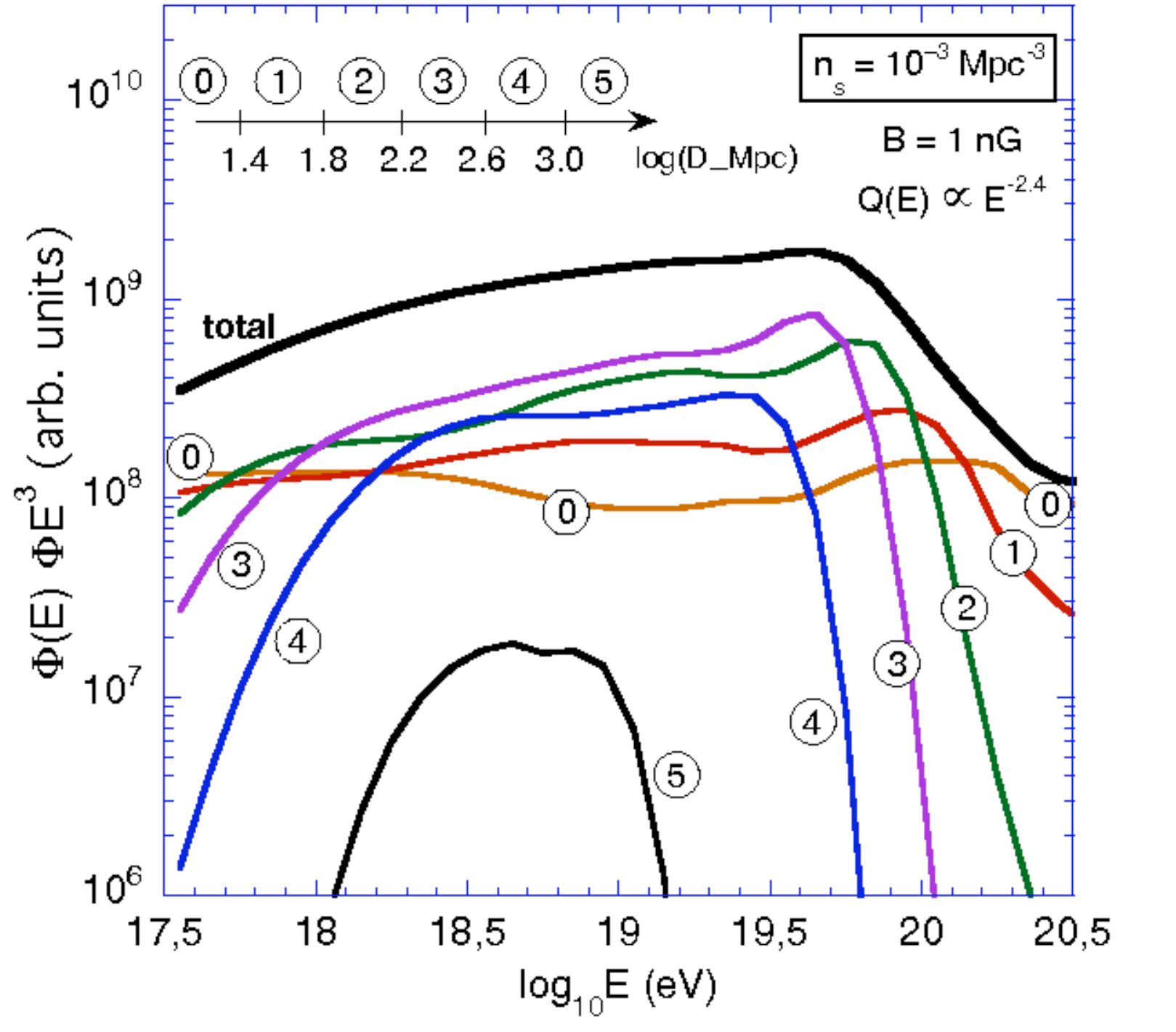}
\hfill \includegraphics[width=0.33\textwidth]{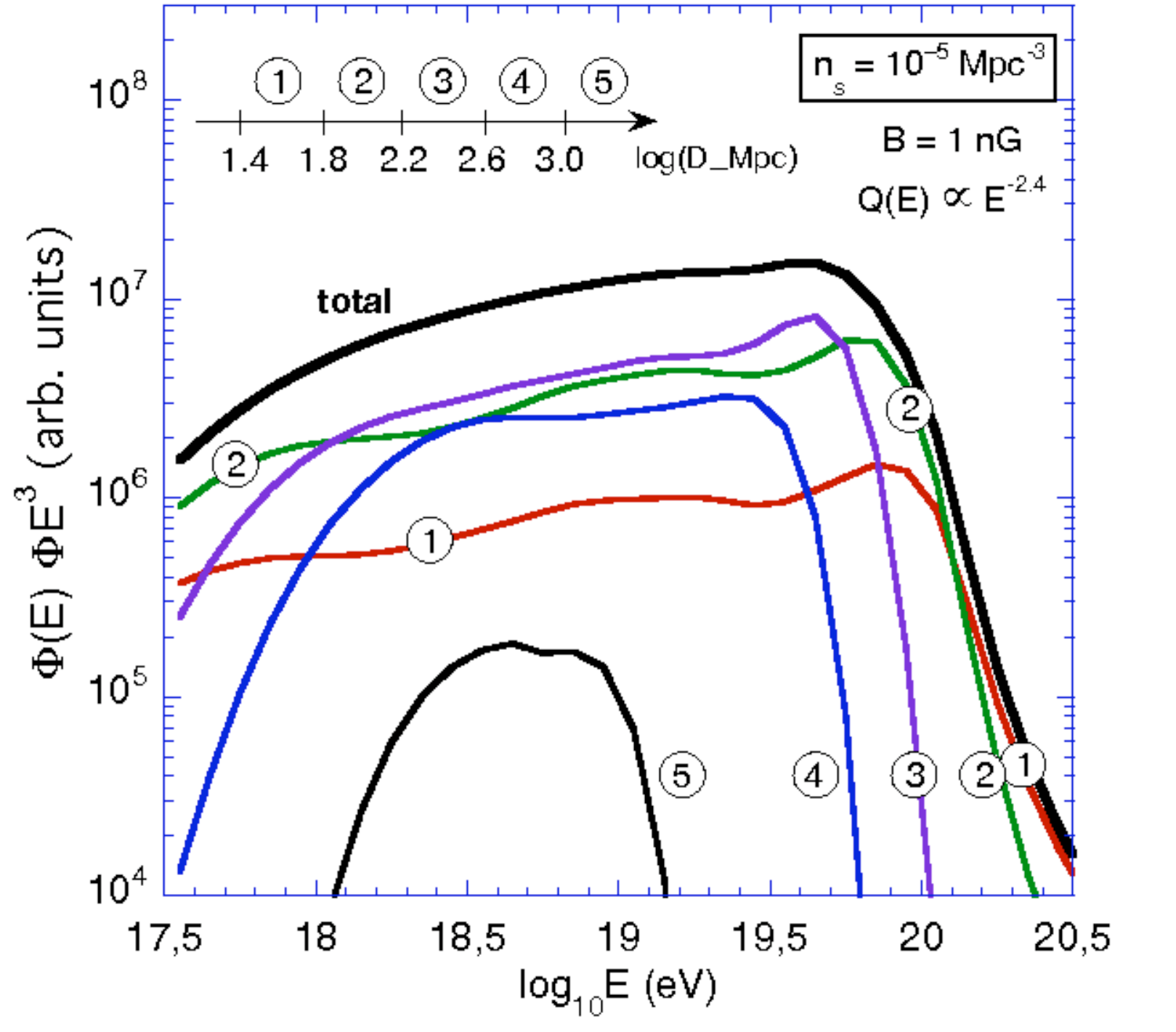}
\caption{Contribution of different intervals of source distances (as indicated by the labels and the ``distance line'') to the overall EGCR spectrum, for a mixed source composition and an injection spectrum in $E^{-2.4}$. Left: continuous distribution of sources (and $B = 0$). Center: $B_{\mathrm{rms}} = 1$~nG, source density $n_{\mathrm{s}} = 10^{-3}\,\mathrm{Mpc}^{-3}$. Right: $B_{\mathrm{rms}} = 1$~nG, source density $n_{\mathrm{s}} = 10^{-5}\,\mathrm{Mpc}^{-3}$.}
\label{fig:ContributionR}
\end{figure*}

\begin{figure*}[t]
\includegraphics[width=0.33\textwidth]{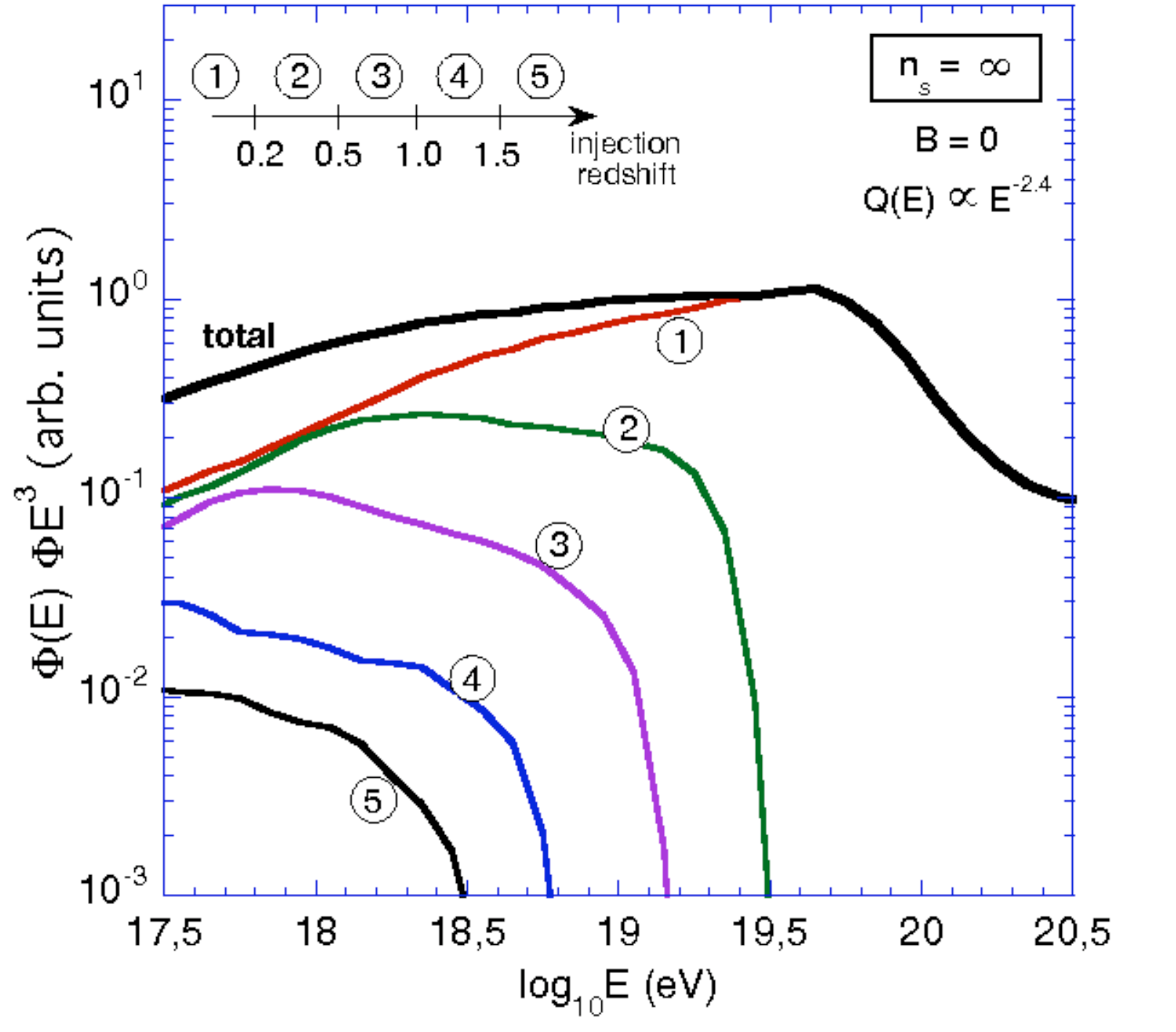}
\hfill \includegraphics[width=0.33\textwidth]{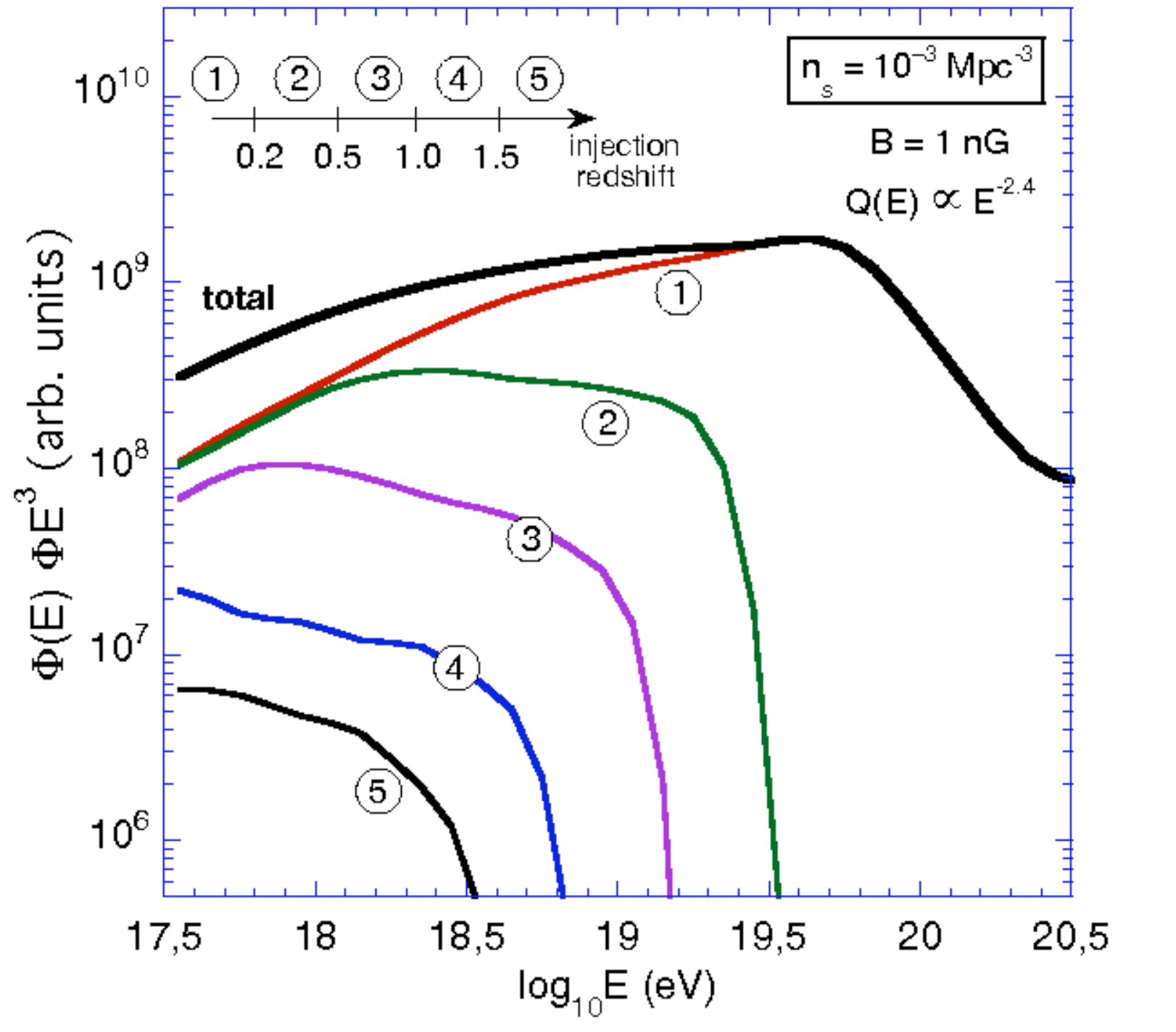}
\hfill \includegraphics[width=0.33\textwidth]{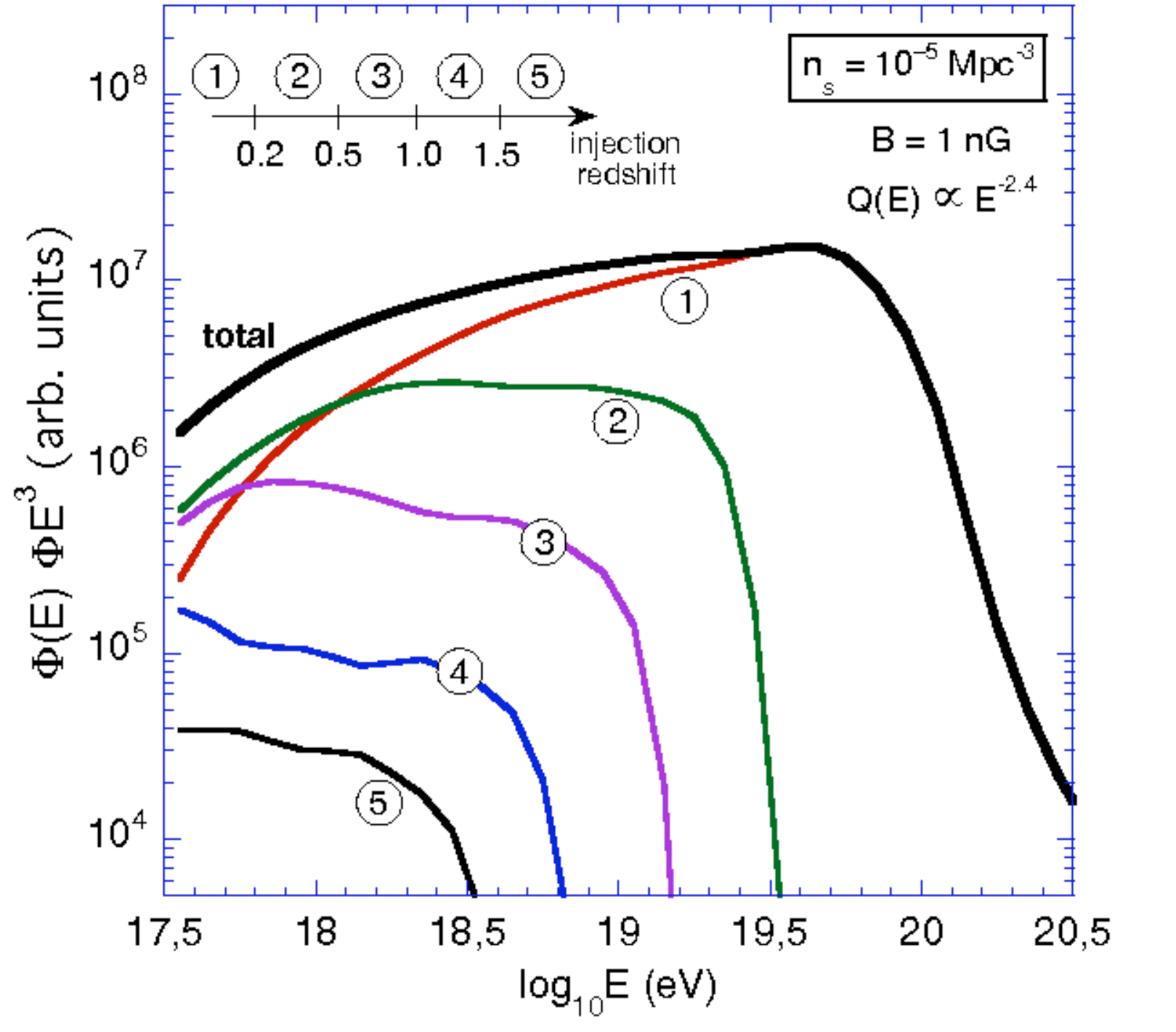}
\caption{Contribution of different intervals of injection redshifts -- i.e. actual times-of-flight -- (as indicated by the labels and the ``injection redshift line'') to the overall EGCR spectrum, for a mixed source composition and an injection spectrum in $E^{-2.4}$. Left: continuous distribution of sources (and $B = 0$). Center: $B_{\mathrm{rms}} = 1$~nG, source density $n_{\mathrm{s}} = 10^{-3}\,\mathrm{Mpc}^{-3}$. Right: $B_{\mathrm{rms}} = 1$~nG, source density $n_{\mathrm{s}} = 10^{-5}\,\mathrm{Mpc}^{-3}$.}
\label{fig:ContributionRed}
\end{figure*}

\subsubsection{Contribution of different source distances}

To better understand the astrophysical origin of the spectral features, we show separately in Fig.~\ref{fig:ContributionR} the contribution of different intervals of source distances, $D$, for the above-mentioned source densities. The intervals chosen define the following ``regions'' of space around the Earth (note that they do not have equal volume nor equal thickness): region~0, for $D\leq 10^{1.4}\,\mathrm{Mpc}$; region~1, for $10^{1.4}\,\mathrm{Mpc}\leq D\leq 10^{1.8}\,\mathrm{Mpc}$; region~2, for $10^{1.8}\,\mathrm{Mpc}\leq D\leq 10^{2.2}\,\mathrm{Mpc}$; region~3, for $10^{2.2}\,\mathrm{Mpc}\leq D\leq 10^{2.6}\,\mathrm{Mpc}$; region~4, for $10^{2.6}\,\mathrm{Mpc}\leq D\leq 10^{3}\,\mathrm{Mpc}$; region~5, for $10^{3}\,\mathrm{Mpc}\leq D$. As can be seen, the contributions to the high-energy flux are very similar for any choice of the source density.

Since the particles propagate almost rectilinearly, the differences in the overall spectrum arise only from the different number of very nearby sources, which leads to a more or less pronounced GZK feature (De Marco et al., 2006). At low energy, however, the spectrum is dominated by the contribution of distant sources when the magnetic field is negligible (due to the large value of $\tau_{\mathrm{90}}$), whereas only nearby sources can contribute significantly if the magnetic field is high enough to settle a diffusive propagation regime and confine low-energy cosmic rays in the vicinity of the sources. In that case, distant sources only contribute at intermediate energies (between $10^{18}$~eV and $10^{19}$~eV), where the two effects of the magnetic field and of energy losses are not strong enough to prevent the particles from reaching the Earth (at least for protons). 

\subsubsection{Contribution of different injection times}

Similarly, we show in Fig.~\ref{fig:ContributionRed} the contribution of different intervals of times-of-flight (i.e. redshifts of injection) to the overall propagated spectrum. As already stressed, intervals of times-of-flight are essentially intervals of distances when the magnetic field is negligible. Otherwise, any given source (at a given distance) can contribute from a large period of injection times, with correspondingly different ``instantaneous propagated spectra''. It appears that the relative contribution of the different intervals are very similar for all source densities, down to low energy. They would actually be totally identical, even in presence of a magnetic field, if the source distribution were continuous. Nevertheless, in the case of discrete sources, even the closest ones cannot contribute at low energy from low injection redshifts (i.e, for short times-of-flight), because of the increased path length imposed by the magnetic field. This explains the observed depletion of the low-redshift contribution at low energy, compared to the case without magnetic field (or with continuous sources), the effect being of course stronger for the lowest source density (or larger values or the magnetic field). The depletion at low redshift is directly visible in the total spectrum and is stronger when the source luminosity is not evolving with time. In the SFR source luminosity evolution case, the lower relative contribution of low redshifts to the total spectrum at low energy in the negligible field case explains the less pronounced depletion in presence of a 1 nG field (see above).

\subsubsection{Contribution of the different nuclei}

\begin{figure*}[t]
\includegraphics[width=0.33\textwidth]{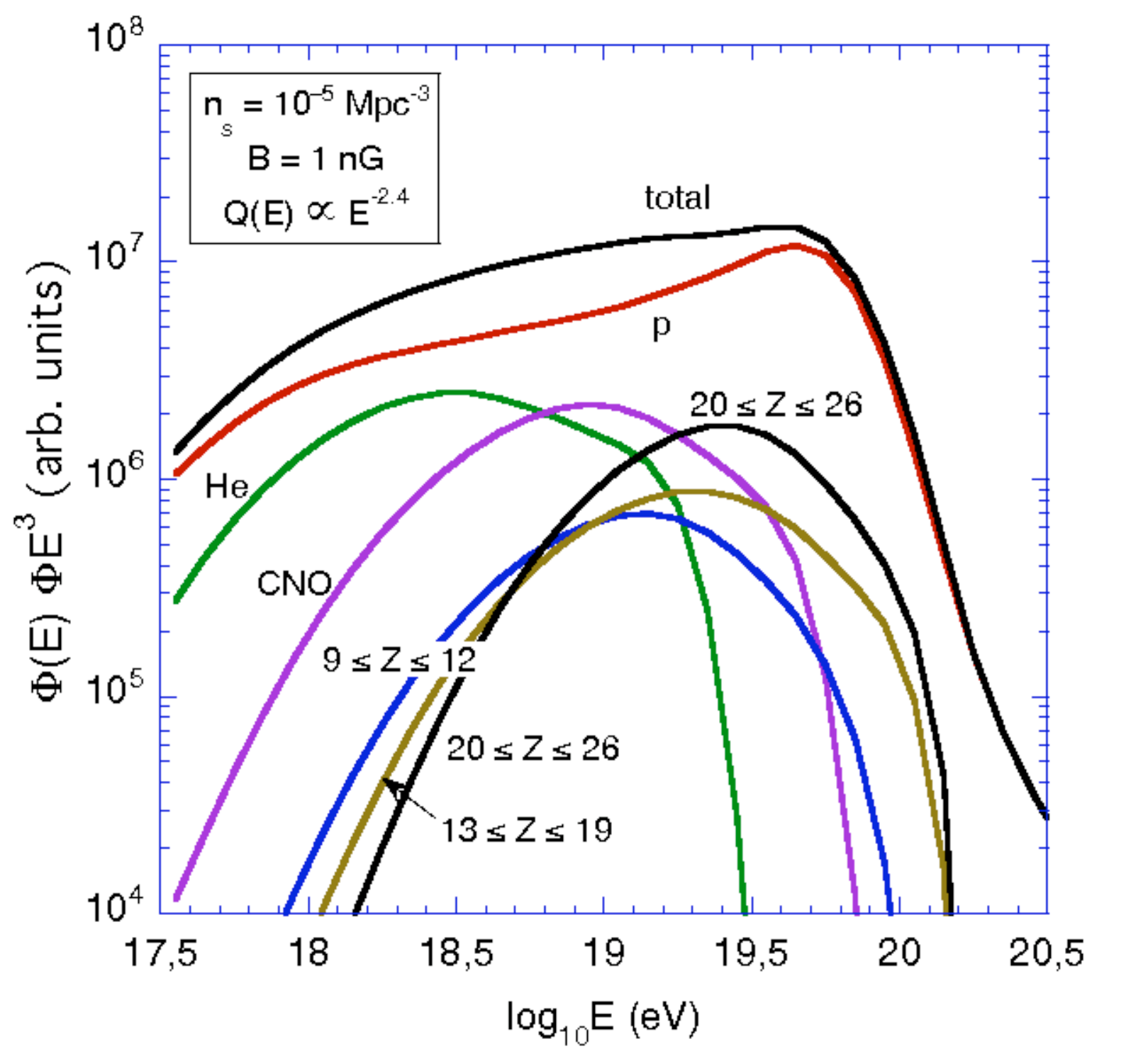} \hfill
\includegraphics[width=0.33\textwidth]{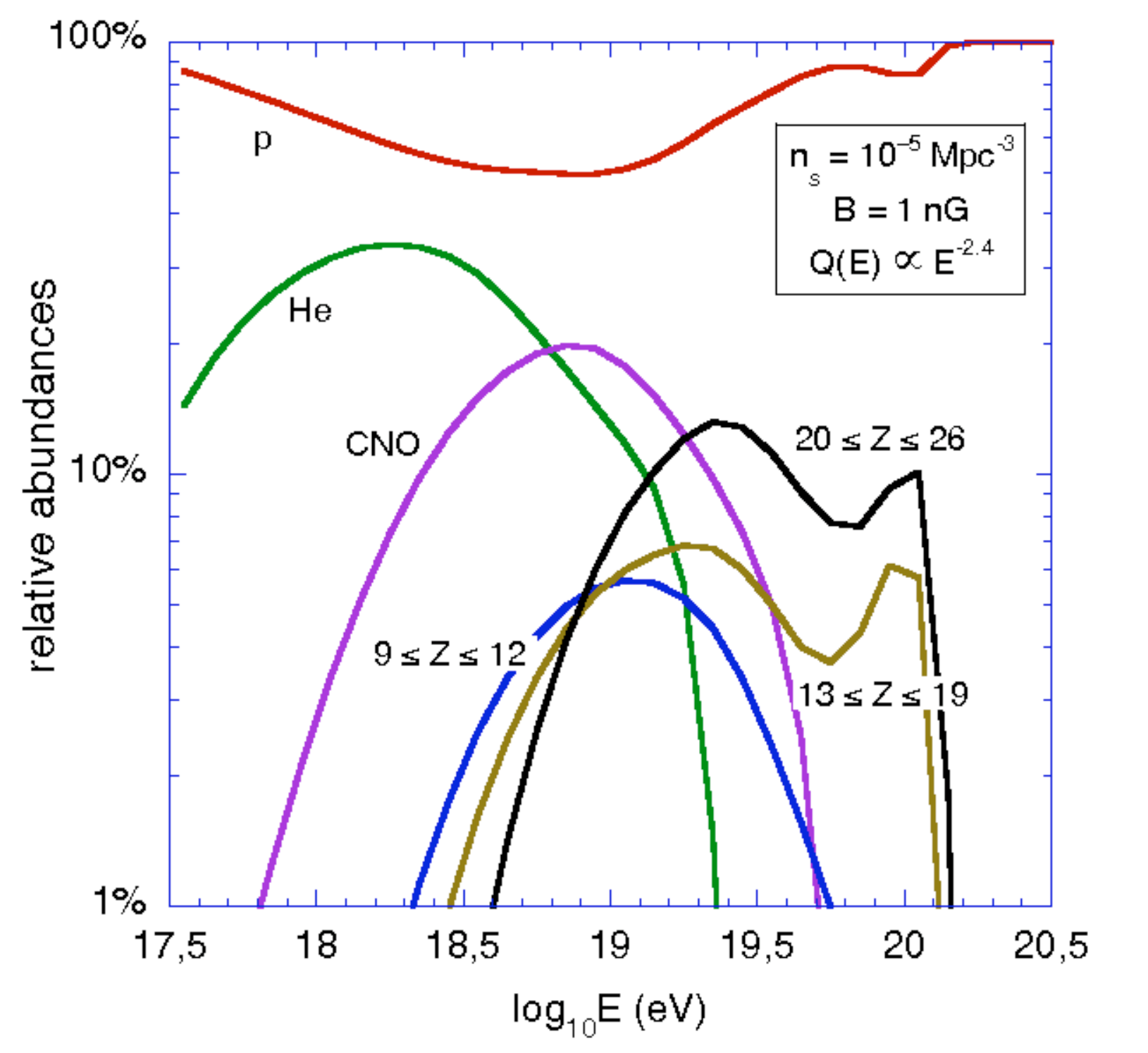}\hfill
\includegraphics[width=0.33\textwidth]{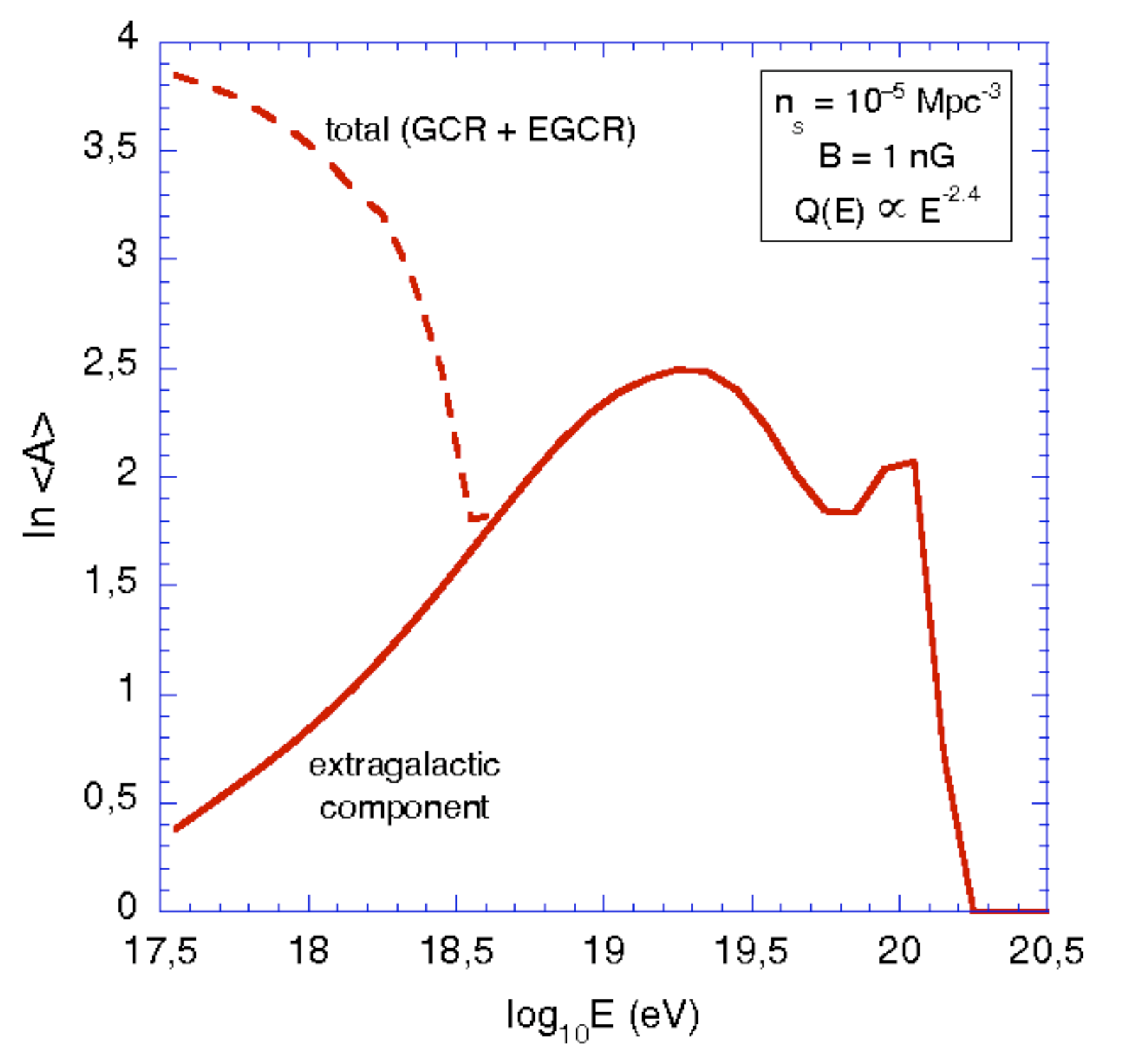}
\caption{Left: contribution of different nuclei to the overall EGCR spectrum, for a mixed source composition (see text), an injection spectrum in $E^{-2.4}$, a homogeneous turbulent magnetic field with $B_{\mathrm{rms}} = 1$~nG, and (one particular realization of) a source density $n_{\mathrm{s}} = 10^{-5}\,\mathrm{Mpc}^{-3}$. Center: corresponding relative abundances as a function of energy (ranges of nuclei are identified through their charge number). Right: logarithm of the corresponding average mass, $\ln\langle A\rangle$, as a function of energy (solid line). Dashed line: resulting value of $\ln\langle A\rangle$ when a pure Fe nuclei Galactic component is added.}
\label{fig:Relat}
\end{figure*}

It is also interesting to study the effect of the EGMF on the different types of nuclei contributing to the propagated spectrum. The relevant quantity being the rigidity, $\propto E/Z$, the effect produced at a given energy is larger for heavier nuclei. As an illustration, we study the case of a homogeneous turbulent magnetic field with $B_{\mathrm{rms}} = 1$~nG and a power-law source spectrum, $Q(E)\propto E^{-2.4}$. Figure~\ref{fig:Relat}a shows the contribution of different mass groups to the overall EGCR spectrum from mixed composition sources, distributed according to a particular Monte-Carlo realization of a source density of $n_{\mathrm{s}}=10^{-5}\,\mathrm{Mpc}^{-3}$. The magnetic horizon effect is clearly observed at increasing energy for increasing nuclei charges. On Fig.~\ref{fig:Relat}b, we show the corresponding relative abundances as a function of energy. As expected, the extragalactic component consists almost exclusively of protons and He nuclei at low energies, where the diffusion coefficient of heavier nuclei is too small to let them reach the Earth in allowed time. As the energy increases, heavier and heavier nuclei can come in, resulting in a larger value of the average mass per cosmic-ray, $\langle A \rangle$, the logarithm of which is shown on Fig.~\ref{fig:Relat}c. The continuous increase of $\ln\langle A \rangle$ up to $\sim 10^{19}$~eV contrasts with the roughly constant (extragalactic) composition expected in this energy range in the case of negligible magnetic fields.

At higher energy, EGCRs get lighter as a result of the progressive photo-disintegration of intermediate-mass and heavy nuclei, starting with the CNO component (He nuclei are photo-disintegrated even earlier, while the heavier components come in). At the highest energies, EGCRs propagate almost rectilinearly and the mass features are independent of the magnetic field. As can be seen, the different energy scales of the GZK-like cutoffs of different nuclei cause the composition to get heavier over a short energy range, between $5\,10^{19}$~eV and $\sim1.5\,10^{20}$~eV (although the upper end of this interval depends on the local density of sources). This composition feature is similar to that discussed in the case of negligible magnetic fields (Allard et al., 2007b).

The solid line of Fig.~\ref{fig:Relat} shows the value of $\ln\langle A \rangle$ for the extragalactic component only. At low energy, a presumably heavy galactic component does contribute significantly to the observed spectrum (up to at least a few EeV in the mixed-composition models), modifying the evolution of the overall $\ln\langle A \rangle$ at the GCR/EGCR transition. This is shown by the dashed line on the same plot.

\subsubsection{Cosmic-ray composition}
\label{sec:CRCompo}

Unfortunately, composition measurements at high energy are difficult and the main observable related to composition that is currently accessible is the average penetration depth of cosmic-ray air showers in the atmosphere. More precisely, when detecting the fluorescence radiation accompanying the development of CR-induced air showers, high-energy cosmic ray observatories can measure the grammage (in g/cm$^{2}$) along the shower axis at which the maximum of development is reached. Although there are some fluctuations from shower to shower (due to the essential stochasticity of the first interactions in the shower development), this grammage -- traditionally referred to as $X_{\max}$ -- is on average an increasing function of energy and a decreasing function of the cosmic-ray mass (i.e. $\langle X_{\max}\rangle$ is smaller for showers induced by Fe nuclei than by protons). It is thus instructing to compute the average value of $X_{\max}$ from the above propagated compositions and compare its energy evolution with the observational data, as proposed in Allard et al. (2005b).

To this aim, an assumption has to be made about the composition of the GCR component, which contributes to the overall CR composition, and even dominates at low energy. Following previous studies and for the sake of simplicity, we assume that the GCRs are composed exclusively of Fe nuclei above $10^{17.5}$~eV, as suggested by the KASCADE results showing a cutoff of lighter nuclei between $10^{15}$~eV and $10^{17}$~eV. Although some intermediate-mass nuclei (like C, N, O or Si) might still be present around $10^{17.5}$~eV, there influence on the results presented here should not be dramatic. For instance, the value of $\langle X_{\max}\rangle$ is typically 30~g/cm$^{2}$ smaller in the case of a pure Fe composition than in the case of a pure CNO composition.

\begin{figure}[t]
\centering{\includegraphics[width=\columnwidth]{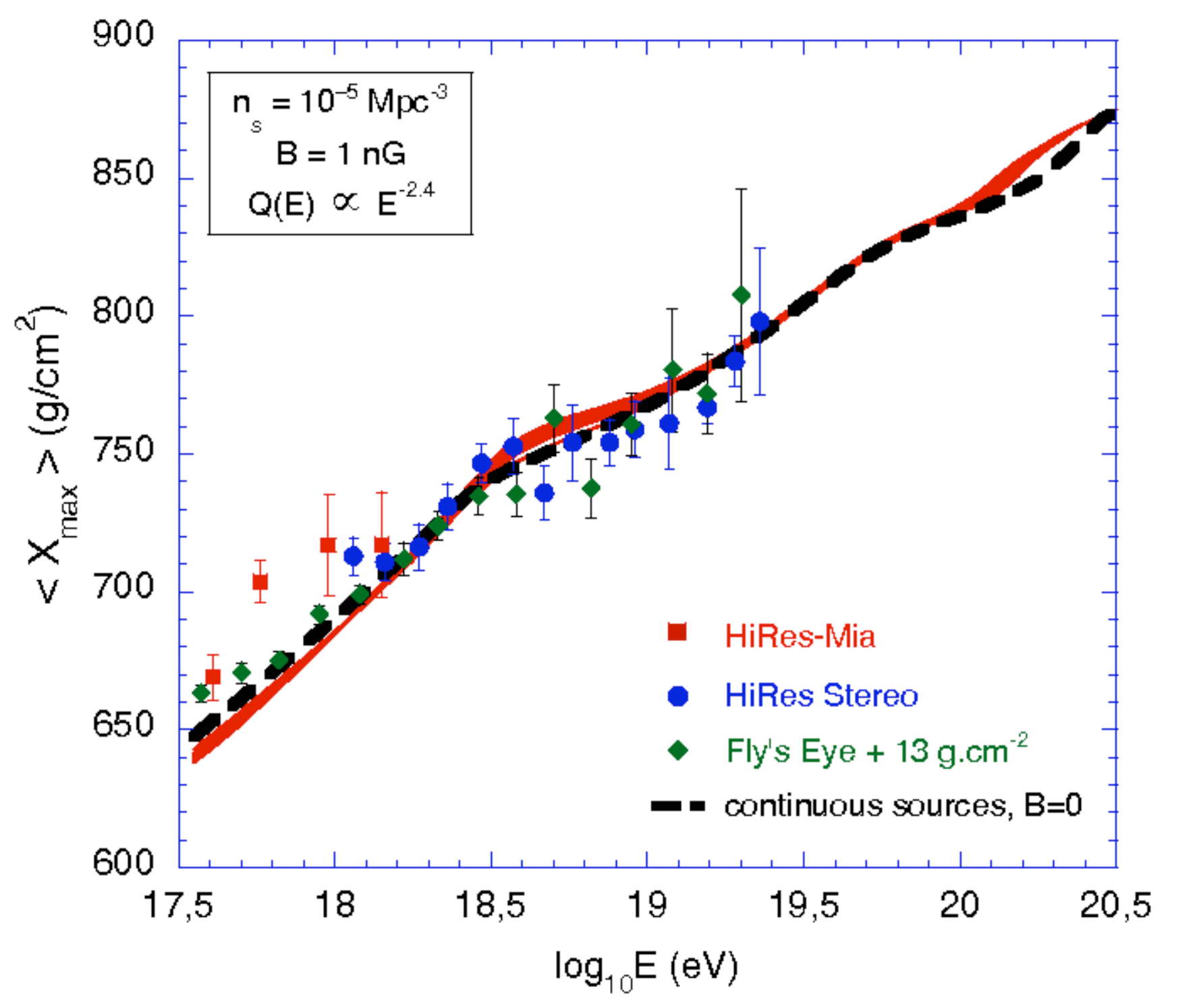}}
\caption{Average atmospheric depth at the maximum of development of the CR showers, $\langle X_{\max} \rangle$, as a function of energy. The curves show the expectation for each of the spectra displayed on Fig.~\ref{fig:mixedCompositionSpectra_density} (the case of a source density $n_{\mathrm{s}} = 10^{-3}\,\mathrm{Mpc}^{-3}$ is omitted for clarity), compared with the HiRes-Mia, Fly's Eye and HiRes Stereo data.}
\label{Erate_1nG-5}
\end{figure}

In Fig.~\ref{Erate_1nG-5}, we show the energy evolution of $\langle X_{\max}\rangle$ in the case of  mixed composition sources with a density $n_{\mathrm{s}} = 10^{-5}\,\mathrm{Mpc}^{-3}$ and a turbulent EGMF with $B_{\mathrm{rms}} = 1$~nG, as compared to the data from HiRes-Mia (Abu-Zayyad et. al.,  2000), Fly's Eye (Bird et al., 1993, rescaled by 13 $\mathrm{g~cm^{-2}}$ upward as suggested by Sokolsky et al., 2005) and HiRes Stereo (Abbasi et al., 2005) experiments. One curve is shown for each EGCR spectrum given in Fig.~\ref{fig:mixedCompositionSpectra_density}: as can be seen, the cosmic variance is rather small when one considers an average quantity such as $\langle X_{\max}\rangle$. The case of a continuous source distribution without magnetic field is also plotted as a reference (dashed line).

The actual value of $X_{\max}$ for a simulated cosmic-ray shower depends somewhat on the underlying hadronic model assumed to compute the high-energy cascade. However, the energy dependence of $X_{\max}$ is similar for all models, so differences come down essentially to a global shift of the curves in Fig.~\ref{Erate_1nG-5}, typically by $\sim 20\,\mathrm{g\,cm^{-2}}$ (see discussions in Allard et al., 2007a,b). Here, we used the QGSJet-II hadronic model (Ostapchenko, 2004), which leads to the best agreement of the mixed composition model with the data.

As can be seen from the comparison with the no-field case, the presence of a 1~nG extragalactic magnetic field does not modify significantly the shape of the $\langle X_{\max}\rangle$ evolution with energy. In particular, they leave unchanged the characteristic features of the mixed composition model to which the data lend support (see Allard et al., 2007a,b). Nevertheless, magnetic fields cause the $\langle X_{\max}\rangle$ evolution to be slightly steeper in the Galactic-to-extragalactic transition region (i.e. between $10^{17.5}$~eV and 3--$4\,10^{18}$~eV), as a direct consequence of the EGCR flux reduction (magnetic horizon effect). The effect on $\langle X_{\max}\rangle$ remains rather small, however, since the EGCR component does not contribute much at low energy (say, below $10^{18}$~eV), even with $B = 0$.

The magnetic suppression of the heavy nuclei in the EGCR component also causes $\langle X_{\max}\rangle$ to be larger when the extragalactic component dominates, i.e. above $\sim 3\,10^{18}$~eV, until the ballistic regime of propagation replaces the diffusive regime (i.e. above $\sim 10^{19}$~eV in this case, for $B = 1$~nG). As heavier nuclei come in, overcoming the magnetic suppression, the EGCR composition gets heavier progressively (see Fig.~\ref{fig:Relat}c), which results in a flatter evolution (smaller increase) of $\langle X_{\max}\rangle$ than in the case without magnetic field. This makes the subsequent inflection point even more pronounced, when the heavy components start falling off (because of the photo-disintegration suffered by nuclei in the extragalactic photon background) and a second transition from heavy to light cosmic rays takes place. Note that the second bump in Fig.~\ref{fig:Relat}c also produces a visible feature in the $\langle X_{\max}\rangle$ evolution plot, although with the same amplitude as in the case without magnetic field, since the propagation is essentially rectilinear at these energies. Finally, the slight difference at the highest energies is an artefact of the continuous source distribution assumed for the reference curve (dashed line). In this energy range, the measured spectrum and composition is directly dependent on the few closest sources in the Earth vicinity, and should be associated with strong anisotropies which leave the study of the overall spectrum (gathering the contribution of all sources) without much purpose.

In sum, we find that the characteristic features of a mixed-composition model -- including a smooth transition from heavy to light cosmic rays across the ankle, an inflection point in the $\langle X_{\max}\rangle$ evolution around $10^{19}$~eV (and a small second bump below $10^{20}$~eV) -- are not attenuated by the extragalactic magnetic fields, but rather accentuated. We furthermore checked that the features  we predict would be even more accentuated by increasing the heavy nuclei relative abundances at the sources without affecting the agreement of the predicted spectrum with data. 

\begin{figure}[t]
\centering{\includegraphics[width=\columnwidth]{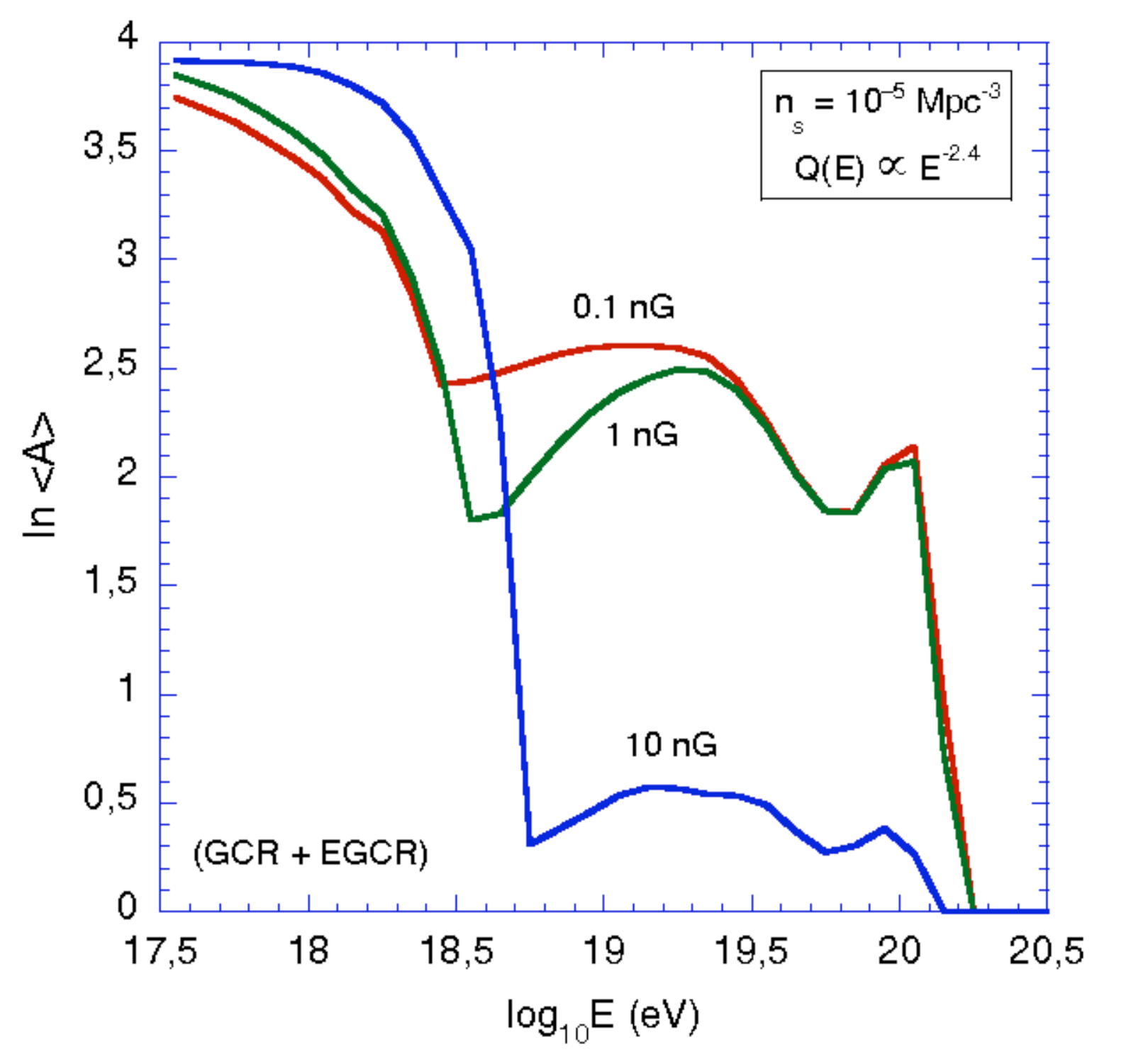}}
\caption{Logarithm of the CR average mass, $\ln\langle A\rangle$, as a function of energy, for different values of the magnetic field (as indicated), a source spectrum in $E^{-2.4}$ and a source density $n_{\mathrm{s}} = 10^{-5}\,\mathrm{Mpc}^{-3}$.}
\label{fig:lnAMoyen_B}
\end{figure}

\begin{figure}[t]
\centering{\includegraphics[width=\columnwidth]{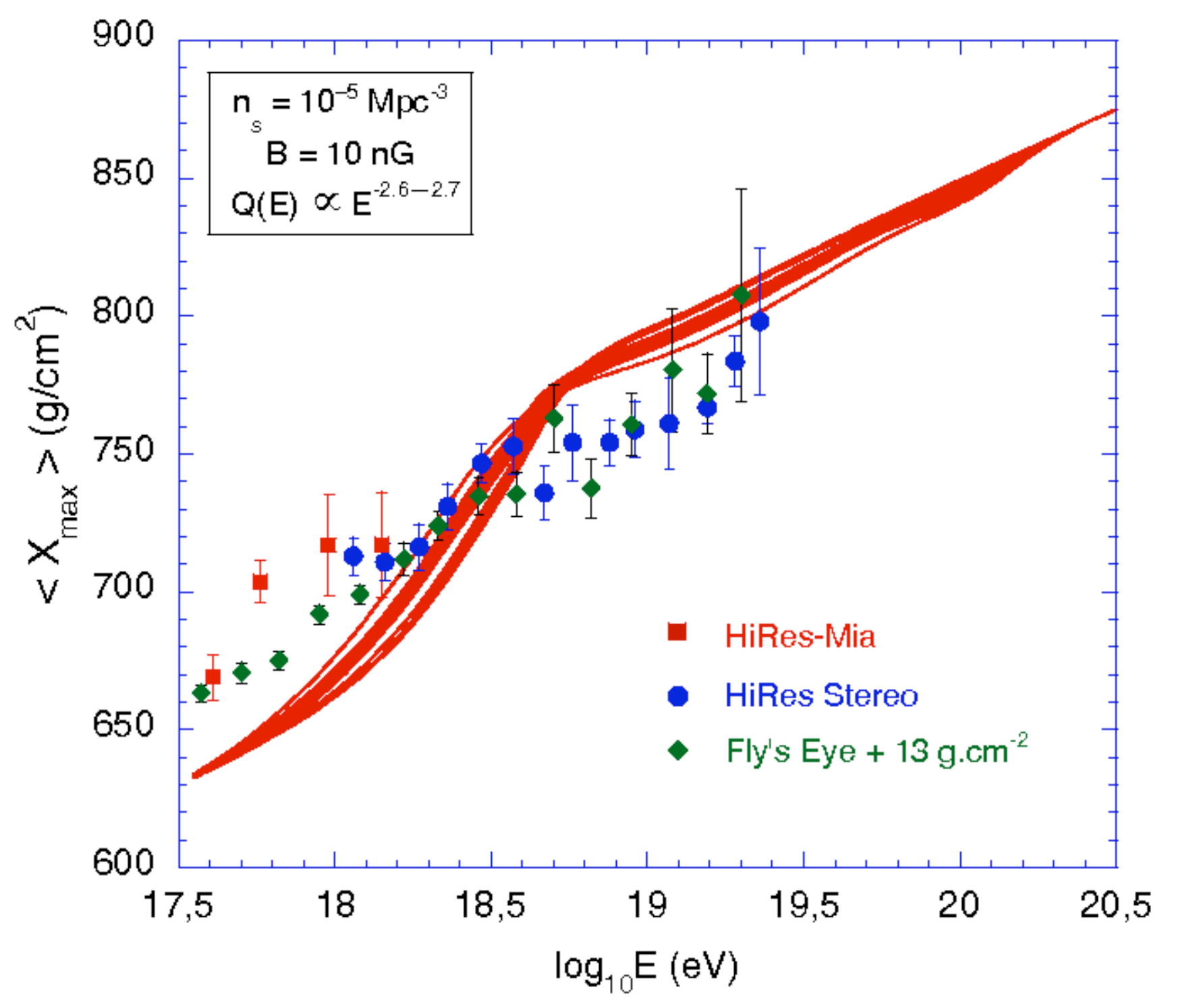}}
\caption{Same as Fig.~\ref{Erate_1nG-5}, for a 10~nG magnetic field.}
\label{fig:Erate_10nG}
\end{figure}

This is a consequence of the dependence of the average CR mass on the intensity of the magnetic field, as shown in Fig.~\ref{fig:lnAMoyen_B}, where we plot $\ln\langle A \rangle$ as a function of energy for three different values of the magnetic field: 0.1~nG, 1~nG and 10~nG. In the latter case, the suppression of the low-energy extragalactic nuclei is so strong (see Fig.~\ref{spectre_field}) that the resulting CR composition sharply drops around the ankle from the Galactic Fe nuclei to an almost pure proton extragalactic component, with only a small fraction of nuclei coming in at higher energy. The resulting evolution of $\langle X_{\max}\rangle$ with energy then appears to be much too steep in this energy range, as illustrated in Fig.~\ref{fig:Erate_10nG}.

\begin{figure}[t]
\centering{\includegraphics[width=\columnwidth]{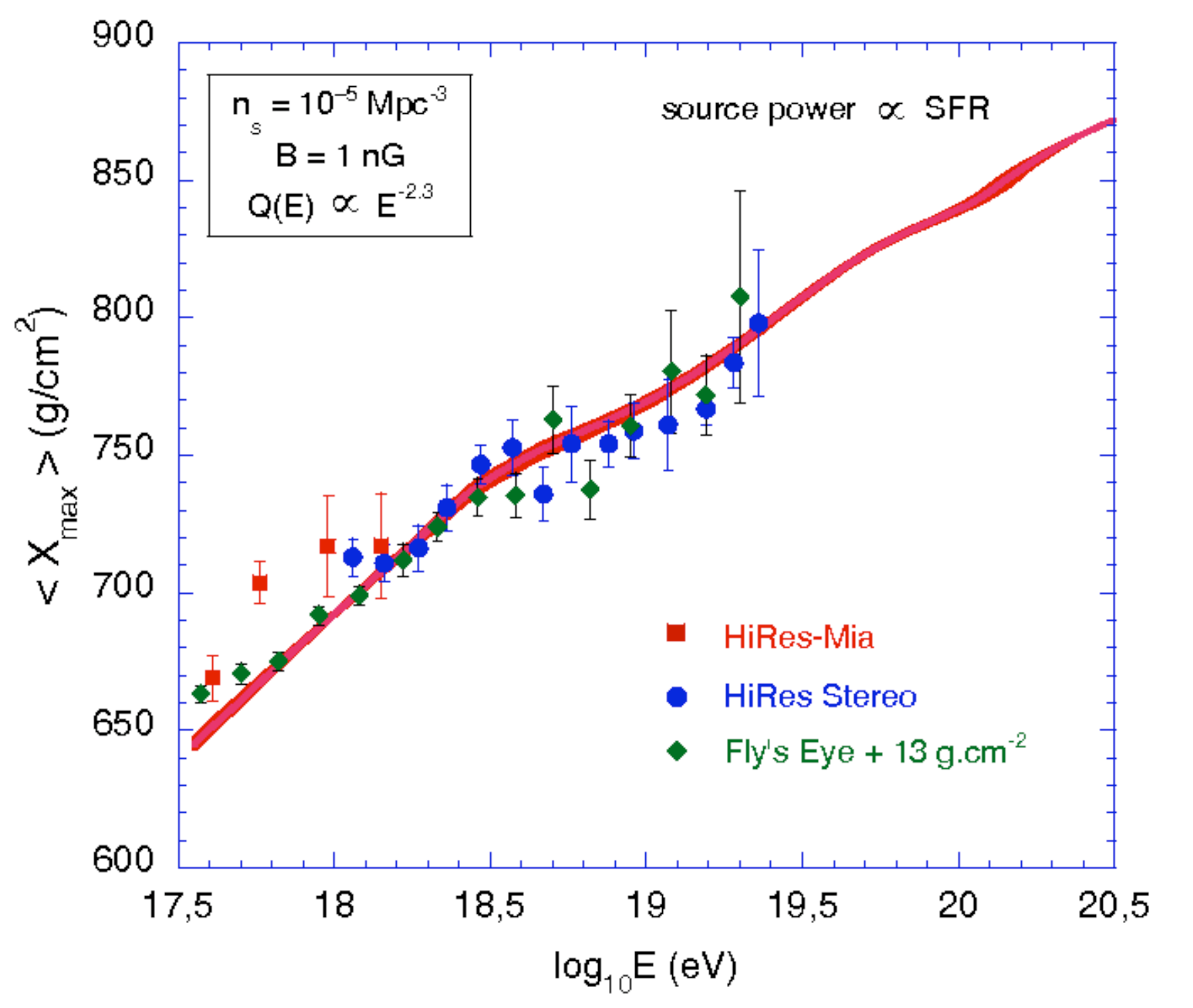}}
\caption{Same as Fig.~\ref{Erate_1nG-5}, for the spectra displayed on Fig.~\ref{spectre_evol} with a source power evolution proportionally to the star formation rate.}
\label{Erate_evol}
\end{figure}

As noted in Sect.~\ref{sec:magnHorizons}, the effect of a magnetic field at low energy can be roughly compensated by an increase of the source power at high redshifts (which most contribute to the low energy flux). The resulting effect on the energy evolution of $\langle X_{\max}\rangle$ is illustrated in Fig.~\ref{Erate_evol}, where larger values of $X_{\max}$ (i.e. a larger fraction of protons, from the EGCR component) is observed at low energy, in better agreement with the data and essentially identical to the case with a steeper source spectrum and $B = 0$ (cf.~Fig.~\ref{Erate_1nG-5}).

\section{Discussion}

Following on the study of the spectral and composition features associated with mixed-composition EGCR source models (Allard et al., 2005; 2007a; 2007b), we investigated here the influence of a non-negligible extragalactic magnetic field and found that the main features of such models remain essentially unchanged as long as the magnetic field intensity is smaller than a few~nG. The astrophysical interpretation of the ankle as a transition from the Galactic to the extragalactic cosmic-ray component remains identical, and relatively strong magnetic fields can even require the GCR component to be extended up to higher energies, into the ankle. Likewise, extragalactic magnetic fields do not alter the characteristic feature in the energy evolution of $X_{\max}$ (the main composition observable in current experiments), namely the flattening of the $X_{\max}$ evolution and the existence of an inflection point between $10^{18.5}$~eV and $10^{19.2}$~eV, in good agreement with the observations. In most cases, it is actually enhanced. For EGMF intensities larger than a few~nG, the resulting energy evolution of $X_{\max}$ becomes incompatible with the data.

The magnetic field model used in the above calculations -- namely, a homogeneous turbulence -- is admittedly simplistic. It is likely that the EGMF shows various levels of structures related to the structures in the matter distribution. However, in the case of external sources, such structures would only lead to smaller propagation effects, since the filling factor of the larger fields would be smaller. Note also that Kotera and Lemoine (2007) considered more realistic margnetic field configurations, with intensities between 0.1 and 10 nG, and found only a weak dependence of the EGCR (pure proton) spectrum above $10^{17}$~eV on the spatial distribution of the field (as controlled by the specific model relating the magnetic field to the matter density). A different conclusion could be reached, however, if most sources were located \emph{inside} highly magnetized regions. In this case, a larger suppression of the low-energy EGCRs could be expected, as recently discussed by Sigl and Armengaud (2005) and Sigl (2007). It should be noted, however, that the influence of a magnetic confinement around the sources on the propagated spectrum remains uncertain, since it greatly depends on the simulation used to compute the magnetic field intensity and its spatial distribution. For instance, the magnetic fields simulated in Dolag et al. (2004) are likely to have a much weaker effect on the charged particles propagation than those of Sigl et al. (2004).

In the absence of any clear prediction regarding their time evolution, we have also assumed constant extragalactic magnetic fields. This should not have a strong impact on our results, however, since we showed on Fig.~\ref{fig:ContributionRed} that most of the particles contributing to the spectrum above $10^{17.5}$~eV were emitted at redshifts smaller than 0.5, i.e. over a period where such an evolution should be quite limited. Even if stronger magnetic fields were present at higher redshifts, the results presented here would remain essentially unchanged. Note that the time evolution model used by Gazizov and Berezinsky (2007) also implies smaller coherence lengths at large redshifts, which partly compensates the effect of a stronger field.

We have also assumed that all EGCR sources are essentially identical, i.e with the same injection spectrum and intrinsic luminosity. This need not be the case, and one may object that sources contributing at the highest energies might be less numerous than lower-energy sources. In this case, the magnetic suppression of low-energy particles would be reduced, and the results closer to those obtained earlier in the case of a negligible magnetic field.

Finally, we stress that the measurement of the UHECR composition is an important step towards the clarification of the source phenomenology as well as the progressive transition from a Galactic to an extragalactic cosmic-ray component. The preliminary Auger data (Unger et al., 2007) appear in good agreement with the Fly's Eye results above $4\,10^{17}$~eV, and thus with the $X_{\max}$ features associated with the mixed composition models, including a possible flattening of the $\langle X_{\max}\rangle$ evolution at the highest energies. It is however too early to determine whether this feature (if any) is associated with the GZK cut-off in the spectrum and whether the composition is getting lighter again, down to protons, above $10^{20}$~eV. Future Auger data, with enhanced statistics and an extension at lower energy (Medina et al., 2006), should be decisive in allowing one to distinguish between the different GCR/EGCR transition models. In addition, as the recent results from KASCADE (KASCADE collaboration, 2007), KASCADE-Grande (KASCADE-Grande collaboration, 2007) and Auger (Engel et al., 2007) have shown, a precise measurement of the properties of the cosmic-ray air showers can help better constrain hadronic models at high-energy and give more confidence in future composition analyses. And since extragalactic magnetic fields may also modify the energy evolution of the CR composition, as was shown above, let us finally note that new constraints on the strength and structure of the EGMF should also result from the analysis of UHECR arrival directions and large statistics anisotropy studies, as expected from the Pierre Auger Observatory and its Northern hemisphere extension (Nitz et al., 2007).

\section*{Acknowledgments}
We thank Angela Olinto, Kumiko Kotera, Martin Lemoine, Gary Mamon and Jean-Christophe Hamilton for helpful discussions.

\end{document}